\documentclass[bj,authoryear]{imsart}

\usepackage{float}
\usepackage{graphicx}
\usepackage{amsmath}
\usepackage{relsize}
\usepackage{amsfonts,amsthm,epsf,stackrel}
\usepackage{bbm}
\usepackage{amssymb}
\usepackage{graphicx,subfigure}
\usepackage{tikz}

\graphicspath{{./pics/}}

\startlocaldefs
\theoremstyle{plain}

\newtheorem{theorem}{Theorem}[section]
\newtheorem{lemma}[theorem]{Lemma}
\newtheorem{algo}[theorem]{Algorithm}
\newtheorem{corollary}[theorem]{Corollary}
\newtheorem{proposition}[theorem]{Proposition}
\theoremstyle{remark}

\newtheorem*{remark}{Remark}

\newcommand{\mE}{\ensuremath{\mathbb E}}
\newcommand{\mP}{\ensuremath{\mathbb P}}
\newcommand{\II}{\ensuremath{\mathbb I}}
\newtheorem*{theorem*}{Theorem}
\newtheorem{assumption}{Assumption}
\newcommand{\CoFilter}{\texttt{CoFilter} }
\endlocaldefs

\begin{document}
\begin{frontmatter}
\title{A procedure for multiple testing of partial conjunction hypotheses based on a hazard rate inequality}
\runtitle{A   procedure for multiple testing of PC hypotheses}

\begin{aug}
\author[A]{\inits{T.}\fnms{Thorsten}~\snm{Dickhaus}\ead[label=e1]{dickhaus@uni-bremen.de}\orcid{0000-0003-3084-3036}}
\author[B]{\inits{R.}\fnms{Ruth}~\snm{Heller}\ead[label=e2]{ruheller@post.tau.ac.il}}
\author[A]{\inits{A.}\fnms{Anh-Tuan}~\snm{Hoang}\ead[label=e3]{anhtuan.hoang@uni-bremen.de}}
\author[C]{\inits{Y.}\fnms{Yosef}~\snm{Rinott}\ead[label=e4]{yosef.rinott@mail.huji.ac.il}}
\address[A]{Institute for Statistics, University of Bremen, Bremen, Germany\printead[presep={,\ }]{e1,e3}}

\address[B]{Department of Statistics and Operations Research, Tel Aviv University, Tel Aviv, Israel\printead[presep={,\ }]{e2}}

\address[C]{Department of Statistics, Center for the Study of Rationality, The Hebrew University of Jerusalem, Jerusalem, Israel\printead[presep={,\ }]{e4}}
\end{aug}

\begin{abstract}
The partial conjunction null hypothesis is tested in order to discover a signal that is present in multiple studies. 
The standard approach of carrying out a multiple test procedure on the  partial conjunction (PC) $p$-values can be extremely conservative. We suggest alleviating this conservativeness, by eliminating many of the conservative PC $p$-values prior to the application of a multiple test procedure. This leads to the following two step procedure: first, select the set with PC $p$-values below a selection threshold; second, within the selected set only,   apply a family-wise error rate or false discovery rate controlling procedure on the conditional PC $p$-values. The conditional PC $p$-values are valid 
if the null p-values are uniform and the combining method is Fisher. The proof of their validity is based on a novel inequality in hazard rate order of partial sums of order statistics which may be of independent interest. We also provide the conditions for which the false discovery rate controlling procedures considered will be below the nominal level.   We  demonstrate the potential usefulness of our novel method, \CoFilter (conditional testing after filtering),  for analyzing multiple genome wide association studies of Crohn's disease. 
\end{abstract}


\end{frontmatter}

\section{Introduction}\label{sec1}

Suppose we have $n$ independent studies, each examining $m\gg 1$ related hypotheses. For example,  in genome wide association studies (GWAS), we can have $n$ independent studies (i.e., $n$ different cohorts), examining each $m\gg 1$   genotypes,  in order to discover the genotypes that are associated with a phenotype \citep{franke2010genome}.  We thus have  $n\times m$   null hypotheses, we call {\em{elementary}} null hypotheses. We also have a  corresponding matrix of $n\times m$ $p$-values, ${\cal P}_{n\times m}$.   The entries of $ {\cal P}_{n\times m}$ are the {\em{elementary}} $p$-values. The  elementary $p$-value for the $j$th elementary hypothesis in the $i$th study is in entry $(i,j)$ of $ {\cal P}_{n\times m}$.  

Consider a set of $n$ elementary null hypotheses $H_{1},\ldots,H_{n}$. For example, in GWAS, $H_i$ is the elementary null hypothesis that there is no association between a particular  genotype and the phenotype in study $i$, for $i=1,\ldots,n$. 
For an integer parameter $1\leq r \leq n$ set by the analyst,  the partial conjunction (PC) null hypothesis is \citep{benjamini2008screening}, 
$$H^{r/n} = \{\text{at most $r-1$ null hypotheses are false}\}.$$
Thus, a rejected PC null hypothesis leads to the conclusion that at least $r$ elementary null hypotheses are false. In GWAS,  for a particular genotype, the rejection of $H^{r/n}$ leads to the conclusion that the genotype is associated with the phenotype in at least $r$ studies. Our interest lies in values $r\in\{2,\ldots,n\}$. In particular,  by setting $r=2$ we can establish minimal replicability: the rejection of  $H^{2/n}$ implies that  the finding is  replicated, in the sense that we rule out that the association of the genotype with the phenotype may exist in at most one study.   

In GWAS as well as in other typical modern applications, the interest lies in testing $m\gg 1$ PC null hypotheses. The aim for GWAS is to discover the genotypes associated with the phenotype in at least $r$ out of the $n$ studies. The different studies may be different cohorts as in \cite{heller2014replicability}, or diverse environments as in \cite{Li21}. 
  Multiple testing of PC null hypotheses has also been useful, in order to discover, for example: the brain voxels in which at least  $r$ out of $n$ covariates are associated with the outcome \citep{benjamini2008screening};   the genes expressed in  at least $r$ out of $n$ time points \citep{Sun11};  the outcomes with at least $r$ out of $n$ evidence factors \citep{Karmakar20}; 
 the features  for which the results  are replicated in at least $r$ out of $n$ studies meta-analyzed (\citealt{bogomolov23} and references within).

Since  each PC null hypothesis is a composite null hypothesis, the test can be very conservative \citep{Dickhaus-JSPI2013, Liang23}. This conservatism can lower the power dramatically when accounting for the multiplicity of testing $m\gg 1$ PC hypotheses. 
The problem is especially acute in the common setting in which all $n$ elementary hypotheses are true for a large proportion of the $m$ PC null hypotheses examined. Next,  in Section \ref{subsec-intro-pc}, we first review typical tests for a single PC null hypothesis,  and  then  the clever solution by \cite{Wang20} for addressing the problem of the tests being conservative when $m\gg 1$.  Their work serves as a backdrop for ours. We conclude the introduction with a brief outline of our contributions in Section \ref{subsec-intro-contributions}.

\subsection{Testing partial conjunction hypotheses}\label{subsec-intro-pc}

Let $p_1,\ldots, p_n$ be mutually independent $p$-values for testing the family of elementary null hypotheses $H_1,\ldots,H_n$.   Denoting by $\leq_{st}$ the usual  stochastic order (first order dominance),  when $H_i$ is true, the $p$-value is {\em valid}, if it satisfies the following stochastic inequalities: $p_i\geq_{st} Uni[0,1]$ (where $Uni[0,1]$ denotes the continuous uniform distribution between zero and one); or equivalently $-\log p_i \leq_{st} Z_i$  for $Z_i\sim Exp(1)$ (where $Exp(1)$ denotes the standard exponential distribution).

 Using the Fisher combination function (Section$~21.1$ in \citealt{fisher1934statistical}), the PC $p$-value is \citep{benjamini2008screening}: 
   \begin{equation}\label{eq-pvfisher}
     p^{r/n} \equiv p^{r/n}(p_{1},\ldots,p_{n}) = 1-F_{\chi_{2(n-r+1)}^{2}}\left(-2\sum_{i=r}^{n}\log\left(p_{(i):n}\right)\right),
   \end{equation}
    where $p_{(1):n}\leq\cdots\leq p_{(n):n}$ are the ordered $p$-values, and  $F_{\chi_{2(n-r+1)}^{2}}$ denotes the cumulative distribution function (cdf) of the chi-square distribution with $2(n-r+1)$ degrees of freedom. 
For testing $H^{r/n}$,  $p^{r/n}$ is a valid $p$-value.   Note that if $H^{r/n}$ is true, at least $n-r+1$ of the $p_i$'s have a distribution that is stochastically at least as large as the uniform. It is easy to see, by a coupling argument, that      $-\sum_{i=r}^{n}\log\left(p_{(i):n}\right)\leq_{st} Z_1+\ldots+Z_{n-r+1}$, where $Z_1,\ldots, Z_{n-r+1}$ are independent identically distributed (iid) $Exp(1)$. 
Therefore, $$\mP\left(p^{r/n}\leq \alpha\right) = \mP\left(-\sum_{i=r}^{n}\log\left(p_{(i):n}\right) \geq q_{1-\alpha}/2\right)\leq   \mP\left(\sum_{i=1}^{{n-r+1}}Z_i\geq q_{1-\alpha}/2\right) = \alpha,$$ where $q_{1-\alpha}$ denotes the $1-\alpha$ quantile of $F_{\chi_{2(n-r+1)}^{2}}$.

The motivation for the PC $p$-value in \eqref{eq-pvfisher} comes from its use in testing global null (or intersection) hypotheses. A  {\em{global null}} hypothesis for $H_{i_1},\ldots,H_{i_{n-r+1}}$, $\{i_1,\ldots,i_{n-r+1}\}\subset\{1,\ldots,n\}$,  is the hypothesis that all the elementary null hypotheses are true (i.e., that $\cap_{k=1}^{n-r+1}H_{i_k}$ is true). Since $H^{r/n}$ is false if and only if the global null hypothesis is false for all possible subsets of size $n-r+1$, the largest of the $\binom{n}{n-r+1}$ global null $p$-values is a valid $p$-value for $H^{r/n}$. 
So we can view the PC $p$-value \eqref{eq-pvfisher} as derived by taking the  Fisher combination function $p^{1/(n-r+1)}:[0,1]^{n-r+1}\rightarrow[0,1]$ for testing the global null on $n-r+1$ hypotheses, that is increasing in each argument, and then defining the PC null hypothesis $p$-value to be the value resulting from the combination of the $n-r+1$ largest $p$-values. Viewing it this way, it is clear that any other combination functions for testing the global null hypothesis can be used instead.  



Broadly, combining functions can be divided into quantile based methods and sum-based methods. An example of a quantile based combining function for the global null is Bonferroni, so a valid test for $H^{r/n}$ is $(n-r+1)p_{(r):n}$. An example of a sum-based method is Fisher. It has the advantage that  even if all $p$-values are only moderately small the evidence against the PC null hypothesis can be large. 

For testing $m\gg 1$ PC null hypotheses, \cite{Wang20}  suggested incorporating a  filtration of the most conservative tests. \cite{Wang20} uses for filtration and testing the Bonferroni combining method for each PC null hypothesis:  the $r-1$ largest $p$-values in order to filter out PC hypotheses, and the $r$th largest $p$-value for  testing (multiplied by $n-r+1$).  See Section \ref{sec-Crohn} for a detailed description of their algorithm for false discovery rate (FDR) control. 
A potential weakness of this approach  is that the combining method is quantile based.  


As in \cite{Wang20}, we consider filtering out large PC $p$-values in order to alleviate the conservativeness. However, we would like to use PC $p$-values computed using the Fisher combining method,  since it has excellent power properties for a wide range of signals \citep{benjamini2008screening, hoang2021combining}.
 Following the filtering step, we would like to apply the multiple testing procedure only on the remaining PC $p$-values (adjusted for the filtering process). 
The simplest, yet potentially very effective, procedure, is to select for multiple testing only the PC $p$-values that are at most as large as a fixed predefined threshold $\tau$. We expect that a far smaller proportion than $\tau$ of the true PC null hypotheses will have PC $p$-values at most $\tau$, due to their conservativeness. 
Thus by selecting all PC $p$-values at most $\tau$, we may greatly reduce the multiplicity problem. 
Of course, the original $p$-values are no longer valid after selection, so adjustment for selection is necessary. 

\subsection{Contributions}\label{subsec-intro-contributions}

Our main methodological contribution is the introduction  in Section \ref{sec2:multiple} of a two-stage multiple test procedure for testing multiple PC hypotheses, called \CoFilter, which first selects  a subset of the $m$ PC hypotheses,   with PC  $p$-values at most a certain threshold $\tau$ (which is either fixed in advanced or estimated from the data). The PC $p$-values get adjusted for the selection event by conditioning on the latter. We  use the Fisher combination test statistic.     
We shall show
using a new hazard rate order result (see \citealt{shaked2007stochastic} Section 1.B for an introduction to the hazard rate order),  that the adjustment for selection is to inflate each selected $p$-value by the factor $1/\tau$. So for the PC hypothesis $H^{r/n}$, the {\em{conditional PC $p$-value}}  is 
 $\frac{1}\tau p^{r/n}$, and the {\em{unconditional PC $p$-value}}  is 
 $p^{r/n}$. 
In the second stage of testing, the conditional PC $p$-values (for which the unconditional PC $p$-values were below $\tau$) are used in a (standard) multiple testing procedure.

Our main theoretical contribution is a novel hazard rate ($hr$) order result, proved in Section \ref{sec-main-theorem}, which is necessary in order to prove that the conditional $p$-values are valid. Denoting $[Q \mid A]$  any random variable whose distribution is the conditional distribution of $Q$ given $A$, we need to show that  $\left[\frac{1}{\tau} p^{r/n} \mid p^{r/n} \leq \tau\right]\geq_{st}Uni[0,1]$ if $H^{r/n}$ is true.  
Let $X_1 = -\log p_1,\ldots, X_n = -\log p_n$, and let $Z_1, \ldots, Z_{n-r+1}$ be iid $Exp(1)$ for $r\in  \{1,\ldots,n\}$. Let $\{X_{(i):n}, i=1, \ldots, n \}$ be the order statistics of the $n$ $X_i$'s.   Then, if $H^{r/n}$ is true, it follows that
\begin{equation}\label{eq-st-PC-sum}
 X_{(1):n}+\ldots X_{(n-r+1):n}\leq_{st} Z_{(1)}+\ldots Z_{(n-r+1)}.
\end{equation}
It is easy to see that stochastic order holds in \eqref{eq-st-PC-sum}  if $X _i \leq_{st} Exp(1)$  for at least $n-r+1$ of the $X_i$’s by a straightforward coupling argument. However, $\left[\frac{1}{\tau}p^{r/n} \mid p^{r/n} \leq \tau\right ]\geq_{st}Uni[0,1]$ if and only if $\leq_{st}$ is replaced with $\leq_{hr}$  in \eqref{eq-st-PC-sum} (a detailed explanation of this claim is presented in the proof of Corollary \ref{cor-main}). For this, we need to assume that $X _i\sim Exp(1)$  for at least $n-r+1$ $X_i$’s, which amounts to assuming simple hypotheses for those $i$’s. For both stochastic and hazard rate orders we need to assume that those $n-r+1$ $X_i$’s or the corresponding  $p_i$’s are independent and that the remaining $r-1$ ones may be dependent but must be independent of the above-mentioned $n-r+1$ $X_i$’s.

  We prove that if at least $n-r+1$ of the $X_i$s are $Exp(1)$, then the $hr$ inequality is satisfied, as formalized in \eqref{eq:mainBBSS}.  We also consider  in Proposition \ref{cor-LehmannAlternatives} the case that all the $X_i$s are $c_i\times Exp(1)$, with  $c_i\geq 0$, and at least $n-r+1$ of $c_i$s are $\le 1$.  

Our procedure, \CoFilter, is particularly suitable for multiple high dimensional studies that examine each  $m\gg 1$  null hypotheses. Within each study, the test statistics are typically dependent. Moreover, the FDR may be the preferred error rate to control since $m$ is large. Therefore, we provide in Section \ref{sec-dependence-FDR} the sufficient conditions for the dependencies among the test statistics within each study, so that a multiple testing procedure targeting FDR control for the family of PC hypotheses,  will indeed control the FDR, at least asymptotically.  

Finally, we apply \CoFilter to show its usefulness in real and simulated data examples. The real data example in Section \ref{sec-Crohn} uses the GWAS data from  \cite{franke2010genome}. They carried out a  meta-analysis of  eight GWAS of Crohn's disease in order to identify loci that are associated with the disease in at least one study. We aim to identify genotypes with association with the phenotype in at least  $r\in\{2,3,4\}$ studies. For Crohn's  disease, this is important for better understanding the disease pathogenesis.



\section{Testing multiple partial conjunction null hypotheses} \label{sec2:multiple}

We consider $m > 1$ (marginal) PC null hypotheses $H^{r/n}_1, \ldots, H^{r/n}_m$  simultaneously. The assumption that $r$ and $n$ are the same for all $1 \leq j \leq m$ is not necessary, and it is only made for notational convenience. 

The $m$ PC null hypotheses are tested using the $n\times m$ matrix of elementary$p$-values (defined in the Section \ref{sec1}), ${\cal P}_{n\times m}$, where column $j$ is used to test $H^{r/n}_j$. 
We shall assume the following for the elementary$p$-values that are used for testing a true PC null hypothesis: 
\begin{assumption}\label{as:for-condpv-validity}
Among the $p$-values $p_1,\ldots,p_n$ for testing a true PC null hypothesis  $H^{r/n}$, there are at least $n-r+1$ that are iid $Uni[0,1]$. 
\end{assumption}

For testing $H^{r/n}_1, \ldots, H^{r/n}_m$, this assumption is satisfied if all the elementary hypotheses are  simple null hypotheses, tested with continuous test statistics,  and the $p$-values across the $n$ rows are independent. This is the case in our example application in Section \ref{sec-Crohn}. We have $n=8$ independent studies, and the elementary hypothesis for a genotype in a study is  that it is not associated with the phenotype (Crohn's disease). The resulting $p$-value for a genotype in a study is $Uni[0,1]$ if the elementary hypothesis is true, and although it depends on other $p$-values in the same study (i.e., in the same row of ${\cal P}_{n\times m}$), it is independent of the $p$-values from other studies (i.e., in other rows).

We shall show in Section \ref{sec-main-theorem} that this assumption is sufficient for the conditional PC $p$-values to be valid, i.e., that for $j$ such that  $H_j^{r/n}$ is true, if $p_j^{r/n} \leq \tau$, then 
 $\left[\frac{1}{\tau}p^{r/n}_j \mid p_j^{r/n} \leq \tau\right]\geq_{st}Uni[0,1]$. This motivates our 
proposal for multiple testing of the $m$ PC hypotheses, called \CoFilter  (conditional testing after filtering).

\begin{algo}[\CoFilter]\label{algo-workflow} $ $
\begin{itemize}
    \item[(i)] Compute for each $j \in \{1, \ldots, m\}$  $p_j^{r/n}$ for testing $H^{r/n}_j$ using \eqref{eq-pvfisher}.
    \item[(ii)] Select all  coordinates $j$ for which $p_j^{r/n} \leq \tau$. Let $S_{\tau} = \{j:  p_j^{r/n} \leq \tau\}$, and let $\frac{1}{\tau}p_j^{r/n}$ be the conditional PC $p$-value for $j\in S_{\tau}$. 
    \item[(iii)] Utilize the conditional $p$-values obtained in step (ii) in a standard multiple test procedure.
\end{itemize}
\end{algo}

Thus, in step (iii), we can apply Bonferroni, or Holm's uniform improvement over Bonferroni \citep{holm1979simple},  at level $\alpha$, and as long as Assumption \ref{as:for-condpv-validity} is satisfied for every true PC hypothesis,  the family-wise error rate (FWER) is controlled at level $\alpha$. For example, for $n$ independent studies each testing $m$ point null hypotheses (with a continuous test statistic), the FWER will be controlled at level $\alpha$, regardless of the dependence between the $p$-values within each study.

Similarly, in step (iii), we can apply the Benjamini-Hochberg (BH) procedure \citep{benhoch1995}  at level $\alpha$. If Assumption \ref{as:for-condpv-validity} is satisfied for every true PC hypothesis, and the PC $p$-values are independent or positive regression dependency on a subset (PRDS, \citealt{Benjamini01}) then the false discovery rate (FDR) is controlled at level $\alpha$. For example, for $n$ independent studies each testing $m$ point null hypotheses (with a continuous test statistic), the FDR will be controlled at level $\alpha$, if the dependence between the $m$ elementary $p$-values within each study is PRDS. We formalize this in Proposition \ref{prop-finitesampleFDR-PRDS}. In 
Section \ref{sec-dependence-FDR} we also provide (asymptotic) justifications for typical high dimensional settings where the elementary $p$-values within each study are locally dependent but not necessarily PRDS.  

 We consider two approaches for selecting $\tau$. Let $S_{\tau}$ be the set of those indices in $\{1, \ldots, m\}$ for which the unconditional $p$-values are not greater than $\tau$.

\begin{description}
\item[$1.$ (Pre-specified $\tau$)] 
Choose a $\tau\in(0,1)$ beforehand. Multiple testing in step (iii) is done on $\left\lbrace \frac{1}{\tau}p^{r/n}_j: j\in S_\tau \right\rbrace$. (Note that the focus on a single $\tau$ is merely for notational convenience, since more generally we can consider a pre-specified vector of $m$ selection thresholds, so that each PC hypothesis has its own pre-specified selection threshold. )
\item[$2.$ (Greedy choice of $\tau$)] 
Choose the value $\tau$ which leads to the largest number of rejections in step (iii) of Algorithm \ref{algo-workflow} when using the  level $\alpha$ BH procedure. 
The choice of this greedy $\tau$ can be written concisely as follows. Recall that the BH cutoff for any family of null hypotheses of size $M$ is $t_{BH} = \max\{t: M\times t\leq \alpha R(t) \}$, where $R(t)$ is the number of $p$-values less than or equal to $t$, and all hypotheses with $p$-values that are at most $t_{BH}$ are rejected. In step (iii) of Algorithm \ref{algo-workflow}, the number of hypotheses is $|S_{\tau}|$, and the number of conditional PC $p$-values less than or equal to $t$ is the number of unconditional PC $p$-values that are at most $\tau t$, i.e., $|S_{\tau t}|$. More formally, for any $t\in [0,1]$, let 
\begin{eqnarray}
&& R(t) = \sum_{i\in S_{\tau}}\II\left( \frac{p_i^{r/n}}{\tau}\leq t \right) = \sum_{i\in S_{\tau}}\II\left( p_i^{r/n}\leq \tau t \right)= \sum_{i=1}^m\II\left( p_i^{r/n}\leq \tau t \right)=|S_{\tau t}|. \nonumber
\end{eqnarray}

Denoting the BH threshold for a given $\tau$ by $\hat t(\tau)$:
$$\hat t(\tau) = \max\left \lbrace  t: |S_\tau|\times t\leq \alpha |S_{\tau t}|\right \rbrace, 
$$ 
we choose 
\begin{equation}\label{greedy-tau}
\hat \tau = \arg\max_{\tau \in\{\tau_1,\ldots,\tau_K \}} S_{\tau \hat t(\tau)},
\end{equation}
where $0<\tau_1< \ldots < \tau_K\leq 1$ is a pre-defined finite set of $K$ candidate values for the selection threshold. In Section \ref{sec-dependence-FDR} we provide a theoretical justification for this greedy choice of $\tau$.

\end{description}

For additional power enhancement, in step (iii), we can  consider using multiple testing procedures that incorporate an estimate of  the proportion $\pi_0(S_{\tau})$ of true PC null hypotheses among the selected ones. In particular, we consider the {\em{adaptive BH procedure}}, which is defined as the level $\alpha/\hat{\pi}_0(S_{\tau})$ BH procedure, where $\hat{\pi}_0(S_{\tau})$ is an  estimate of $\pi_0(S_{\tau})$.  The specific estimate we  consider  for the fraction of true null hypotheses in $S_{\tau}$  is 
\begin{equation}\label{eq-pi0hat}
\hat \pi_0(S_{\tau}) =\frac{1+\left\lvert\left\lbrace j: \frac{1}{\tau}p^{r/n}_j>0.5, j\in S_{\tau}\right\rbrace\right\rvert}{(1-0.5)|S_{\tau}|},
\end{equation} which is (a slight variation on) the plug-in estimator of \cite{Schweder82} for the fraction of true PC null hypotheses following selection (i.e., among all hypotheses with PC $p$-value at most $\tau$). 
If all $n\times m$ elementary $p$-values are  independent, and $p$-values from elementary null hypotheses that are true are uniformly distributed, the adaptive BH procedure on $\left\lbrace \frac{1}{\tau}p^{r/n}_j: j\in S_\tau \right\rbrace$ guarantees control of the FDR. This follows from the general theorem in \citealt{Storey03} which states that the adaptive BH procedure with  the (upwardly biased) estimator in \eqref{eq-pi0hat} controls the FDR when the test statistics are mutually independent. More realistically, if the $m$ elementary $p$-values in each row of ${\cal P}_{n\times m}$ are dependent,  in Section \ref{sec-dependence-FDR} we provide theoretical guarantees for asymptotic FDR control using the adaptive BH procedure in step (iii) of \CoFilter.

Our motivation for suggesting the use of adaptive procedures comes from the fact that they are especially useful when the fraction of null hypotheses is small.   Even if in the family of $m$ PC null hypotheses considered,  the fraction of PC null hypotheses is close to one, following selection the fraction of true PC null hypotheses among the selected may be far smaller than one.


\section{Main theorem and proof}\label{sec-main-theorem}

We say that the random variable $X$ and $Y$ satisfy $X \le_{hr} Y$ (\textit{hr=hazard rate order}) if $\mP(Y>t)/\mP(X>t)$ is non-decreasing in $t$. 

\begin{theorem}\label{thm:expord} Let $X_1, \ldots, X_{\ell}$ and $Z_1, \ldots, Z_{\ell}$ be iid $Exp(1)$ variables. For some $n>\ell$ let $X_{\ell+1}, \ldots, X_{n}$  be any further $n-\ell$ positive random variables, which may be dependent, but are independent of $X_1, \ldots, X_{\ell}$. Then
	\begin{equation}\label{eq:mainBBSS} X_{(1):n}+\cdots+X_{(\ell):n} \le_{hr} Z_1 \cdots+ Z_\ell. 
	\end{equation}
\end{theorem}

See Section \ref{subsec-mainproof} for the proof. We note that 
the same proof shows that we can generalize Theorem  \ref{thm:expord} for the case that  $X_i \sim$ Gamma$(\alpha_i, \theta)$ for $i=1,\ldots,\ell$ and $X_{\ell+1}, \ldots, X_{n}$ any further $n-\ell$ positive random variables independent of $X_1, \ldots, X_{\ell}$,  and conclude that $$X_{(1):n}+\cdots+X_{(\ell):n} \le_{hr} \text{Gamma}(\alpha, \theta) \,\,\text{for any}\,\, \alpha\ge \sum_{i=1}^\ell \alpha_i.$$
Theorem  \ref{thm:expord} is the special case that $X_i \sim$ Gamma$(\alpha_i=1, \theta=1)$, and $\alpha = \ell$.


\begin{corollary}\label{cor-main}
For a true PC null hypothesis $H^{r/n}$,  let $p_1,\ldots, p_n$ satisfy that  at least $n-r+1$ of the $p_i$'s are iid $Uni[0,1]$ (i.e., Assumption \ref{as:for-condpv-validity} holds). Then for the PC $p$-value $p^{r/n}$ in \eqref{eq-pvfisher}, the conditional PC $p$-value $\left[\frac{1}{\tau} p^{r/n}\mid p^{r/n}\leq \tau\right]$ is valid for every fixed $\tau\in (0,1]$.  
\end{corollary}
\begin{proof}
By definition, $\left[\frac{1}{\tau} p^{r/n}\mid p^{r/n}\leq \tau\right]$ is valid for testing $H^{r/n}$ if and only if, when $H^{r/n}$ is true,  
\begin{equation}\label{eq-cond-validity}
\frac{\mP(p^{r/n}\leq t)}{\mP(p^{r/n}\leq \tau)}\leq \frac{t}{\tau} \ \forall \ t\in[0, \tau],\end{equation} which is satisfied if and only if  for $X_i = -\log p_i, i\in \{1,\ldots,n\}$, when $H^{r/n}$ is true,
$$ \frac{\mP\left( 2\sum_{i=1}^{n-r+1} X_{(i):n}\geq  \chi^2_{1-t,2(n-r+1)} \right)}{\mP\left( 2\sum_{i=1}^{n-r+1} X_{(i):n}\geq  \chi^2_{1-\tau,2(n-r+1)} \right)}\leq  \frac{\mP\left( 2\sum_{i=1}^{n-r+1} Z_i\geq  \chi^2_{1-t,2(n-r+1)} \right)}{\mP\left( 2\sum_{i=1}^{n-r+1} Z_i\geq  \chi^2_{1-\tau,2(n-r+1)} \right)}, $$ where the RHS is $t/\tau$. Put differently, for validity of the conditional $p$-value we need to show that  $\frac{\mP\left( 2\sum_{i=1}^{n-r+1} X_{(i):n}\geq  x \right)}{\mP\left( 2\sum_{i=1}^{n-r+1} Z_i\geq  x \right)}$ is increasing in $x$, which follows from Theorem \ref{thm:expord}  with $\ell = n-r+1. $
\end{proof}

 Corollary  \ref{cor-main} provides the justification for using the conditional PC $p$-values \eqref{eq-pvfisher} in  \texttt{CoFilter}, when elementary $p$-values from true null hypotheses are $Uni[0,1]$. Following the proof in Section \ref{subsec-mainproof}, we provide in Section \ref{appendix-LehmannAlternatives} an additional result to justify the use of \eqref{eq-pvfisher} for the case that elementary $p$-values  come from the family of specific distributions called Lehmann's alternatives  \citep{Lehmann53} . In Section \ref{sec6} we discuss extensions to cases when elementary $p$-values from true null hypotheses have more general (stochastically larger than uniform) distributions. 

\subsection{Proof of Theorem \ref{thm:expord}}\label{subsec-mainproof}
\begin{proof}
	Set $X_{\ell+1}=Y_1, \ldots, X_{n}=Y_{n-\ell}$. If $\ell>n-\ell$, set $Y_{(n-\ell+1)}=\ldots = Y_{(\ell)} = \infty.$ , and let $Y_{(i)}$ denote the order statistics of the $\max\{n-\ell, \ell\}$ Y's.
	Then $X_{(1):n}+\cdots+X_{(\ell):n} = X_{(1):\ell}\wedge Y_{(\ell)}+\cdots+X_{(\ell):\ell}\wedge Y_{(1)}$, where $X_{(i):n}$ and  $X_{(i):\ell}$ denote  the $i$th order statistic among $X_1, \ldots, X_{n}$ and $X_1, \ldots, X_{\ell}$, respectively. The relation \eqref{eq:mainBBSS} is equivalent to 
		\begin{equation}\label{eq:casem<BBSS}
		X_{(1):\ell}\wedge Y_{(\ell)}+X_{(2):\ell}\wedge Y_{(\ell-1)}+\cdots+X_{(\ell):\ell}\wedge Y_{(1)}\le_{hr} Z_1 \cdots+ Z_{\ell}.
	\end{equation}
	Since the $hr$ order is closed under mixtures of one  side of the of the $\le_{hr}$ relation, in order to prove \eqref{eq:casem<BBSS} it suffices to prove that for any positive constants $K_1,\ldots,K_{\max\{n-\ell, \ell\}}$, possibly = $\infty$,
	\begin{equation}\label{eq:casegenBBSS}X_{(1):\ell}\wedge K_{(\ell)}+X_{(2):\ell}\wedge K_{(\ell-1)}+\cdots+ X_{(\ell):\ell}\wedge K_{(1)}\le_{hr} Z_1 \cdots+ Z_{\ell} ,\end{equation}
	and then take expectation over $K_{(1)},\ldots,K_{(\ell)}$ with respect to the joint distribution of $Y_{(1)}, \ldots, Y_{(\ell)}$, which amounts to conditioning and unconditioning. 

	Set $S=X_1+ \ldots +X_{\ell}$, $L_p=X_{(\ell)}+X_{(\ell-1)}\ldots+X_{(\ell-p+1)}$,  $M_p=K_{(1)}+\ldots+K_{(p)}$, and $R_p=1-L_p/S$, $p=1,\ldots,\ell$. The LHS of \eqref{eq:casegenBBSS} is equal to $T$ defined by 
	\begin{equation*}\label{eq:TTT}
		T:=	\min\{S,\, S-L_p+M_p, \,p=1,\ldots,\ell\}=\min\{S,\, R_pS+M_p, \,p=1,\ldots,\ell\}.
	\end{equation*}
	Then the claim of \eqref{eq:casegenBBSS} is equivalent to  
	\begin{equation}\label{eq:YYYR}
		T <_{hr} S,\,\, \text{that is,}\,\, \,\,\frac{\mP(S>t)}{\mP(T>t)} \,\,\text{is nondecreasing in}\,\,\, t \ge 0.
	\end{equation}
	Note that $R_1,\ldots,R_\ell$ are independent of $S$ because the vector
	$(X_1,\ldots,X_{\ell})/S$ is independent of $S$ for exponential variables. Therefore it suffices to consider \textit{constants} $R_1,\ldots,R_{\ell}$, satisfying $R_1 \ge R_2 \ge \cdots \ge R_{\ell}$ and then integrate over their joint distribution to obtain the result. 	Then $T$ is a function of $S$ only, and we may use $T(S)$.
	 Also, the constants $M_p$ satisfy  $M_1\le M_2 \le \cdots \le M_{\ell}$. Because of this monotonicity, it is easy to see that there exist constants $0\le s_1 \le \cdots \le s_{\ell} \le s_{\ell+1}=\infty$ such that $T=S$ for $S\le s_1 =R_1s_1+M_1$,\,  $T=R_pS+M_p$ for $s_p \le S \le s_{p+1}$,  and $R_ps_{p+1}+M_p=R_{p+1}s_{p+1}+M_{p+1}$, $p=1,\ldots,\ell-1$. See Figure \ref{fig-main-pf}.
	\begin{figure}
		\centering
		\begin{tikzpicture}[domain=0:6]
			\draw[->] (-0.2,0) -- (6.2,0) node[right] {$S$};
			\draw[->] (0,0) -- (0,5.5) node[above] {};
			
			\draw[thick,->,color=red] (0,0) -- (5.5,5.5) node[above] {\tiny {$\bf T_0(S)=S$}};
			
			\draw[thick,->,color=blue] (0,0.5)node at (-0.2,0.5) {\tiny$M_1$} -- (5.5,3.2) node[above] {\tiny 
				{$\bf T_1(S)=R_1S+M_1$}};
			
			\draw[thick,->,color=black] (0,1.1)node at (-0.2, 1.15) {\tiny$M_2$} -- (5.5,2.8) node[below] {\tiny $\qquad \qquad\qquad {\bf T_2(S)=R_2S+M_2}$};
			
			\draw[thin, blue, fill=black] (0.92, 0) circle(0.4mm)node[below] {\tiny $s_1$};
			
			
			\draw[thin, blue, fill=black] (3.15, 0) circle(0.4mm)node[below] {\tiny $s_2$};
			
			\draw[thick, ->,color=purple] (0,.9)node at (-0.15,.9) {{\tiny $t_1$}} -- (0.9,.9) node[above] {};
			
			\draw[ ->, color=purple] (0,2.1)node at (-0.15,2.1) {{\tiny $t_2$}} -- (3.1,2.1) node[above] {};
		\end{tikzpicture}
		\caption{For $\ell=2$, the three lines $T_0(S)=S$,  $T_1(S)=R_1S+M_1$, and $T_2(S)=R_2S+M_2$. The function $T(S)$ is the minimum of the three, that is, their lower envelope.}
		\label{fig-main-pf}
	\end{figure}

	First consider the range $t \le s_1=R_1s_1+M_1 :=t_1$.
	$T(S)$ is an increasing function of $S$, and it is easy to see that $S \ge T$ ($T$ being a minimum over terms including $S$).
	If $S < s_1$ then $T(S)=S$ and therefore for $t\le s_1$  we have that $S\ge t$ implies $T(S)\ge T(t) =t$. It follows that  for $t\le s_1$ we have $S \ge t$ if and only if $T \ge t$ and hence $\displaystyle \frac{\mP(S>t)}{\mP(T>t)}=1$.

	Next, consider the range of $t$ satisfying $t_1=R_1s_1+M_1 \le t \le R_2s_2+M_2:=t_2$.
	Define the random variables $T_1:=[R_1S+M_1 \mid {S>s_1}]$ and 
	$S^*:= [S \mid {S>t_1}]$, where $[Q \mid A]$  denotes any random variable whose distribution is the conditional distribution of $Q$ given $A$. Since  $R_1s_1+M_1=t_1$,  the support of both $T_1$ and $S^*$ is $[t_1, \infty)$.  For $t$ satisfying $R_1s_1+M_1 \le t \le R_2s_2+M_2$ we have $T>t$ if and only if $T_1>t$ and we want to show that in this range the ratio
	$\displaystyle\frac{\mP(S^*>t)}{\mP(T_1>t)}$ is increasing for $t\ge t_1$, that is, $T_1 \le_{hr}S^*$. This together with the fact that the ratio is obviously $\ge 1$ will imply that $\displaystyle\frac{\mP(S>t)}{\mP(T>t)}$ is increasing for $t \le R_2s_2+M_2=t_2$.
	
	In order to show that $T_1 \le_{hr}S^*$ we use the following part of Theorem 1 from \cite{Yu09}.
	
	\begin{theorem*}[Yu, 2009] \label{thm:exp} Let the random variables $X$ and $Y$ have densities $f(x$) and $g(x)$ respectively, both supported
		on $(0, \infty)$. Assume the log density ratio $l(x) = \log(f(x)/g(x))$ is continuous and moreover
		concave, i.e., $X \le_{lc} Y$. Then $X \le_{st} Y$ and $X \le_{hr} Y$ are equivalent.
	\end{theorem*}
	It is obvious that $T_1 \le_{st}S^*$. The density of $S^*$ is proportional to $g(s)=s^{n-1}e^{-s}$ and the density $f(s)$  of $T_1$ is proportional to $g(\frac{s-M_1}{R_1})$ and a straightforward differentiation shows that $\frac{d^2}{ds^2}\log\frac{f(s)}{g(s)}<0$, proving log-concavity.
	
	Before going to the general case we  consider the next range, that is, $t$ satisfying $t_2=R_2s_2+M_2 \le t \le R_3s_3+M_3$ and now define $S^*:=[S\mid S>t_2]$ and $T_2:=[R_2S+M_2\mid S>s_2]$. Since ${\mP(T_1>R_2s_2+M_2)}={\mP(T_2>R_2s_2+M_2)}$   and $S^*=S$ for $S>R_2s_2+M_2$ we have 
	$\displaystyle\frac{\mP(S>t )}{\mP(T_1>t)}=\displaystyle\frac{\mP(S^*>t)}{\mP(T_2>t)}$ for $t=R_2s_2+M_2=t_2$,\,
	and we have to show that the latter ratio is increasing for $t>t_2$. The variables $S^*$ and $T_2$ have a common support $[t_2, \infty)$. The fact that $T_2 \le_{hr}S^*$ follows exactly as the previous case of $T_1 \le_{hr}S^*$ by Theorem \ref{thm:exp}.

	In general, consider the range where $t$ satisfies  $t_p:=R_ps_p+M_p \le t \le R_{p+1}s_{p+1}+M_{p+1}:=t_{p+1}$.
	Now define $S^*:=[S\mid S>t_p]$ and $T_p:=[R_pS+M_p\mid S>s_p]$.
	Since ${\mP(T_{p-1}>t_p)}={\mP(T_p>t_p)}$  and $S^*=S$ for $S>t$ for $t$ in this range, we have 
	$\displaystyle\frac{\mP(S>t )}{\mP(T_{p-1}>t)}=\displaystyle\frac{\mP(S^*>t)}{\mP(T_p>t)}$ for $t = t_p$
	and we have to show that the latter ratio is increasing for $t>t_p$. The variables $S^*$ and $T_p$ have a common support $[t_p, \infty)$ and again $T_p \le_{hr}S^*$ follows from Theorem \ref{thm:exp} as above. This completes the proof.
	
\end{proof}

\subsection{A result for non-uniform null p-values}\label{appendix-LehmannAlternatives}

Suppose the $n$ elementary $p$-values have each a distribution from the family of Lehmann alternatives, introduced  in \cite{Lehmann53}. The CDF of the $i$th $p$-value is 
\begin{equation}\label{eq-Lhemann}
\mP(P_i\leq p_i) = p_i^{d_i} \ \ \textrm{with} \ \ d_i>0.    
\end{equation}
 If  $d_i>1$ then the $p$-value is stochastically larger than uniform.

\begin{proposition}\label{cor-LehmannAlternatives}
Assume the  $p$-values are independent and satisfy each \eqref{eq-Lhemann}.  Then  for the PC $p$-value $p^{r/n}$ in \eqref{eq-pvfisher}, the conditional PC $p$-value $\left[\frac{1}{\tau} p^{r/n}\mid p^{r/n}\leq \tau\right]$ is valid for every fixed $\tau \in (0,1]$.  
\end{proposition}

\begin{proof} 
If $H^{r/n}$ is true, then $d_i\ge 1$ for at least $n-r+1$ $p$-values coming from true elementary null hypotheses. For  $X_i=-log p_i$, $i=1,\ldots,n$, note that  $X_i\stackrel{d}= c_i Z_i$, where $c_i = 1/d_i$ and $Z_i\sim Exp(1)$. Since  $\left[\frac{1}{\tau} p^{r/n}\mid p^{r/n}\leq \tau\right]\geq_{st} Uni[0,1]$ if and only if  $\sum_{i=1}^{n-r+1} X_{(i): n}\leq_{hr} \sum_{i=1}^{n-r+1} Z_i$, it is enough to show that the hazard rate order is satisfied if at least $n-r+1$ of the $c_i$'s are $\leq 1$. This follows directly from Theorem \ref{thm-forLehmann} below. 
\end{proof}
\begin{theorem}\label{thm-forLehmann}
Suppose $X_i \stackrel{d} = c_i Z_i$, $i=1,\dots,n$, where $c_i\geq 0 \ \forall i$,   
 $0\leq c_i \le 1$, for $ i= 1, \ldots, n-r+1$,  and $Z_i\sim Exp(1)$ are independent. Then $\sum_{i=1}^{n-r+1} X_{(i): n}\leq_{hr} \sum_{i=1}^{n-r+1} Z_i$. 
\end{theorem}
\begin{proof}

We shall use the following result in \cite{Nagaraja2006} for independent  $X_i \sim Exp(\lambda_i)$, $i=1,\ldots,n$.  Here $Exp(\lambda)$ has density $\lambda e^{-\lambda t}, \,t>0$. Let 
$\{D(i) = \ell\} = \{X_{(i)} = X_{\ell}\}, 1 \le i, \ell \le n$, that is, $D(i)$ is the index of the $i$th order statistic, which is a random variable. Then
\begin{equation}\label{eq:Nag2} 
	\left\lbrace X_{(i)}, 1 \le i \le n \right\rbrace \overset{d}{=} \left\lbrace \frac{Z_1}{\lambda_{D(1)}+ \ldots + \lambda_{D(n)}}+\ldots +
\frac{Z_i}{\lambda_{D(i)}+ \ldots + \lambda_{D(n)}}, \,1 \le i \le n\right\rbrace
\end{equation}
where and $Z_i \sim Exp(1)$  are iid and independent of the anti-rank
vector $(D(1),\ldots, D(n))$.

We are in the case that $X_i \sim Exp(\lambda_i)$, where $\lambda_i = 1/c_i:= d_i$, with $d_i\geq 1$ for $i=1, \ldots, n-r+1$. 
From \eqref{eq:Nag2}, it follows that $\sum_{i=1}^{n-r+1} X_{(i):n}$  has the following distribution: 
$$\sum_{i=1}^{n-r+1} \sum_{j=1}^i \frac{Z_j}{\lambda_{D(j)}+ \ldots + \lambda_{D(n)}} = \sum_{i=1}^{n-r+1}\frac{(n-r+2-i)Z_i}{\lambda_{D(i)}+ \ldots + \lambda_{D(n)}}.$$
The random variable $W_i: = \frac{(n-r+2-i)}{\lambda_{D(i)}+ \ldots + \lambda_{D(n)}}$ is positive for $i\in \{1, \ldots, n-r+1  \} $, and bounded above by one, since there are at least $n-r+1 - (i-1) = n-r+2-i$ rates $d_j$ that are $\ge 1$ in the denominator's sum. 
Note that for $c\le 1$ and $Z\sim Exp(1)$ we have $cZ\le_{hr} Z$ because $\overline F_Z(t)/\overline F_{cZ}(t)=e^{-t}/e^{-t/c}= e^{t(1/c-1)}$ which 
is increasing (or use Theorem 1.B.21 in \citealt{shaked2007stochastic}). Therefore, $[W_iZ_i \mid D(1)=d(1), \ldots, D(n) =d(n)]\leq_{hr} Z_i$,  where $d(1),\ldots,d(n)$ is a permutation of $1, \ldots, n$. 
By Theorem 1.B.4 in \cite{shaked2007stochastic}, it follows that $[\sum_{i=1}^{n-r+1}W_iZ_i \mid D(1)=d(1), \ldots, D(n) =d(n)] \leq_{hr} \sum_{i=1}^{n-r+1} Z_i$. 

Since this hazard rate order is satisfied for  any given $D(1), \ldots, D(n)$, it follows by Theorem 1.B.8 in \cite{shaked2007stochastic} that the relation is maintained also unconditionally, i.e.,  $\sum_{i=1}^{n-r+1}W_iZ_i \leq_{hr} \sum_{i=1}^{n-r+1} Z_i$, thus completing the proof.  \end{proof}



\section{FDR control under dependence across PC \texorpdfstring{$p$}{p}-values}\label{sec-dependence-FDR}
For high dimensional studies, the test statistics within each study are typically dependent. Moreover, the FDR is the standard criterion for controlling false positive findings in such high-dimensional settings.  We expect our approach with the target of FDR control to be highly robust against  dependencies within each study, for the following reasons.  First, a vast amount of empirical evidence suggests that the (adaptive) BH procedure controls the FDR at the nominal level for most dependencies occurring in practice, and this robustness carries over to our novel approach which applies the (adaptive) BH procedure on conditional PC $p$-values. 
Second, the dependency among PC $p$-values is less severe than within individual studies. As $r$ increases, the PC $p$-values are less dependent, since the overlap between the identity of studies  combined to form the PC $p$-values is reduced (and the studies are independent). 
Third, we have theoretical (finite and asymptotic) FDR guarantees under some additional assumptions. For positive regression dependency on a subset (PRDS, cf. \cite{Benjamini01})  within each study the finite sample FDR is controlled for a fixed value of $\tau$, as we show next in  Proposition \ref{prop-finitesampleFDR-PRDS}. For local dependency within each study, the asymptotic FDR is controlled, as we show in   Proposition \ref{propo-BH-fixed-tau-asymptotic-FDR} for the case of a fixed $\tau$ and in Corollary \ref{justification-greedy-tau} for the greedy choice $\hat{\tau}$ introduced in \eqref{greedy-tau}.


We start by considering \texttt{CoFilter}  using a pre-specified $\tau$ and the BH procedure in step (iii).  We show that  the FDR is controlled at the nominal level if the original within study $p$-values are  positive regression dependent on the subset of true null hypotheses.   

\begin{proposition}\label{prop-finitesampleFDR-PRDS}
For each $j \in \{1, \ldots, m\}$, if $H^{r/n}_j$ is true, 
assume that $\left[\frac{1}{\tau}p^{r/n}_j \vert j \in S_{\tau}\right]$ is a valid (conditional) $p$-value (this is satisfied if the conditions of Corollary \ref{cor-main} or Proposition \ref{cor-LehmannAlternatives} are met). Moreover, assume that 
for each study (i.e., for each row in ${\cal P}_{n\times m}$) the $p$-values are PRDS on the subset of $p$-values corresponding to true null hypotheses, and the $p$-values across studies are independent (i.e,  $p$-values in different rows, in ${\cal P}_{n\times m}$, are independent). In addition, assume that the PRDS property is preserved for every subset of $p$-values  under marginalization over the remaining $p$-values\footnote{ This is satisfied under the subset pivotality condition 2.1 in \cite{Westfall93}, e.g., when we have one-sided $p$-values for the normal means problem
.}. Then the FDR of  the BH procedure at level $\alpha$ on $\left\lbrace \frac{1}{\tau}p^{r/n}_j: j\in S_\tau \right\rbrace$, for a fixed pre-specified $\tau$, is at most $\alpha$.
\end{proposition}

\begin{proof}
According to Theorem 4.1 in \cite{Bogomolov21}, applying the level $\alpha$ BH procedure to the  PC $p$-values $\{p^{r/n}_j, j=1,\ldots,m \}$ guarantees that the FDR on the PC null hypotheses is controlled at level $\alpha$ if in ${\cal P}_{n\times m}$, the rows are PRDS, and for each column that corresponds to a true $H^{r/n}_j$, there are at least $n-r+1$ independent elementary $p$-values that correspond  to $n-r+1$ true elementary hypotheses. 
Since the PRDS property is preserved for any subset of columns of  ${\cal P}_{n\times m}$,  the level $\alpha$ BH procedure on $\left\lbrace \frac{1}{\tau}p^{r/n}_j: j\in S_\tau \right\rbrace$ controls the FDR for  the family of null hypotheses $\{H_j^{r/n}, j\in S_{\tau} \}$. Specifically, denoting by $V_m^{BH}$ and $R_m^{BH}$ the number of true and total PC hypotheses rejected by the \texttt{CoFilter} procedure with a pre-specified $\tau$ and level $\alpha$ BH at step (iii), we guarantee that  $$\mE\left(\frac{V_m^{BH}}{R_m^{BH}\vee 1}\mid S_{\tau} \right)\leq \alpha.$$ By taking the expectation of the above conditional FDR, the  FDR guarantee $\mE\left(\frac{V_m^{BH}}{R_m^{BH}\vee 1} \right)\leq \alpha$ follows  for  $\{H_j^{r/n}, j=1,\ldots,m \}$. 
\end{proof}

\begin{remark}
    The independence assumption in Proposition \ref{prop-finitesampleFDR-PRDS} can be relaxed as follows. In ${\cal P}_{n\times m}$, for each column that corresponds to a true $H^{r/n}_j$, there are at least $n-r+1$ independent elementary $p$-values that correspond  to $n-r+1$ true elementary hypotheses. 
\end{remark}

In the remainder of this section, we analyze asymptotic ($m \to \infty$) properties of multiple tests operating on the proposed conditional PC $p$-values. To this end, we make use of the weak dependence concept introduced by  \cite{Storey03}. They define  'weak dependence' as any type of dependence where   \eqref{eq-conv1}  below holds. They make use of the following  three conditions \eqref{eq-conv1} - \eqref{eq-conv3} in order to provide asymptotic FDR control.
\begin{equation}
\forall t\in (0,1]: \lim_{m\rightarrow \infty}\frac{V_m(t)}{m_0} = G_0(t) \textrm{ and } \lim_{m\rightarrow \infty}\frac{R_m(t)-V_m(t)}{m-m_0} = G_1(t)  \textrm{ a.s.},  \label{eq-conv1}
\end{equation}
where $m_0$ is the number of true null hypotheses; $V_m(t)$ is the (random) number of true null hypotheses with $p$-values below $t$; 
 $R_m(t)$ is the (random) number of  $p$-values below $t$;  and $G_0$ and $G_1$ are continuous functions such that 
\begin{equation}\label{eq-conv2}
\forall t\in (0,1]: 0<G_0(t)\leq t; 
\end{equation}
\begin{equation}\label{eq-conv3}
\lim_{m\rightarrow \infty} \frac{m_0}{m} = \pi_0 \textrm{ exists}.
\end{equation} 
For a pre-defined fixed  tuning parameter $\lambda\in (0,1)$, they define  $\hat \pi_0(\lambda) = (m-R_m(\lambda)) / [(1-\lambda)m]$ to be the plug-in estimate (upwardly biased) for the proportion of true null hypotheses.
In their Theorem 6, \cite{Storey03} prove that if the convergence assumptions in \eqref{eq-conv1}--\eqref{eq-conv3} hold, then for each $\delta>0$, $$\lim_{m\rightarrow \infty} \inf_{t\geq \delta}\left\{ \frac{\hat \pi_0(\lambda) t}{\{R_m(t)\lor 1\}/m}- \frac{V_m(t)}{R_m(t)\lor 1} \right\} \geq 0 $$
holds almost surely. 
The implication of this result is that asymptotically, the false discovery proportion (FDP, defined as $V_m / (R_m\vee 1)$),  is almost surely at most $\alpha$ if the adaptive BH procedure at level $\alpha$ is applied on the $m$ $p$-values (so the FDR, which is the expected FDP, is also controlled at level $\alpha$ asymptotically.) The reason is, that the decision rule of the adaptive BH procedure implies that the threshold $t = t_\alpha$ for the $p$-values is such that (the observable quantity) 
$\hat{\pi}_0(\lambda) t / (\{R_m(t)\lor 1\}/m)$ is upper-bounded by $\alpha$. 

In many high dimensional applications, the dependence within the study (i.e, within each row of the $n\times m $ matrix of elementary $p$-values) is local and hence weak, thus it is reasonable to assume the aforementioned convergence assumptions. This is the case in chief genomics applications: for GWAS or eQTLs, the covariance structure is that of a banded matrix; for microarrays or RNA-seq experiments, gene-gene networks are typically sparse \citep{Wang20}. 

We next provide guarantees of our methodology, when inferring on  multiple studies with weak dependence.  
For the family of PC hypotheses, using the (adaptive) BH procedure on the PC $p$-values  provides asymptotic FDP (and FDR) control if the convergence assumptions \eqref{eq-conv1}--\eqref{eq-conv2} are satisfied for the (joint distribution of  the) PC $p$-values. This is guaranteed, for example, if within each study the dependence is banded or block diagonal, as detailed in Corollary \ref{coro-weak-dependence} below. (Assumption \eqref{eq-conv3} is primarily for notational and mathematical convenience, but not essential, for the results below.) 


Our algorithm, in which we first select the PC hypotheses which have PC $p$-values at most the selection threshold $\tau$, and then applies  the (adaptive) BH procedure on the conditional PC $p$-values, also  provides asymptotic FDP (and FDR) control under weak dependence, as the following proposition asserts, the proof of which is deferred to the appendix. 

\begin{proposition}\label{propo-BH-fixed-tau-asymptotic-FDR}
For each $j \in \{1, \ldots, m\}$, if $H^{r/n}_j$ is true, 
assume that $\left[\frac{1}{\tau}p^{r/n}_j \vert j \in S_{\tau}\right]$ is a valid (conditional) $p$-value (this is satisfied if the conditions of Corollary \ref{cor-main} or  Proposition \ref{cor-LehmannAlternatives} are met). Moreover, assume that the convergence assumptions of equations \eqref{eq-conv1}--\eqref{eq-conv3} hold for the PC $p$-values $\{p_j^{r/n}, j=1,\ldots,m\}$. Then asymptotically almost surely, for a fixed pre-specified $\tau>0$,
(i)  the FDP of the  BH procedure at level $\alpha$ on $\left\lbrace \frac{1}{\tau}p^{r/n}_j: j\in S_\tau \right\rbrace$ is  at most $\alpha$.
(ii) the FDP of the  adaptive BH procedure at level $\alpha$ on $\left\lbrace \frac{1}{\tau}p^{r/n}_j: j\in S_\tau \right\rbrace$ is  at most $\alpha$.
\end{proposition}

Our next goal is to generalize Proposition \ref{propo-BH-fixed-tau-asymptotic-FDR} to cases in which the value of $\tau$ is selected on the basis of the available data. Thus, in Proposition \ref{propo-BH-chosen-tau-asymptotic-FDR} below we consider a random variable $\hat{\tau}$ which describes the selection rule, meaning that $\tau = \hat{\tau}(\text{data})$. We denote by $\pi_1 = 1-\pi_0$ the limiting fraction of non-null PC hypotheses. 

\begin{proposition}\label{propo-BH-chosen-tau-asymptotic-FDR}
For each $j \in \{1, \ldots, m\}$, if $H^{r/n}_j$ is true, 
assume that $\left[\frac{1}{\tau}p^{r/n}_j \vert j \in S_{\tau}\right]$ is a valid (conditional) $p$-value for every fixed $\tau\in (0,1]$ (this is satisfied if the conditions of Corollary \ref{cor-main} or Proposition \ref{cor-LehmannAlternatives} are met). Moreover, assume that the convergence assumptions of Equations \eqref{eq-conv1}--\eqref{eq-conv3} hold for the PC $p$-values $\{p_j^{r/n}, j=1,\ldots,m\}$. Furthermore, assume that the limiting ecdf $
\pi_0 G_0(x) + \pi_1 G_1(x)$ has a unique crossing point $x^*(\alpha) \in (0, 1)$ with the "Simes line" $
x / \alpha$. 

For any given value $\tau \in (0, 1]$, define the (random) function $\hat{F}_\tau : [0, 1] \to [0, 1]$ by 
$\hat{F}_\tau(x) = |S_{\tau x }| / |S_{\tau}|$. Let $\hat{\tau} \equiv \hat{\tau}_m$ denote any $(0, 1]$-valued random variable which is measurable with respect to (the $\sigma$-field generated by) the available (random) data. Assume that the sequence $\left\{\hat{\tau}_m\right\}_{m \geq 1}$ possesses an almost sure limiting value $\tau_\infty \in (0, 1]$, and that
$\vert\vert \hat{F}_{\hat{\tau}_m} - \hat{F}_{\tau_\infty} \vert\vert_\infty \to 0$ almost surely as $m \to \infty$.

Then, the FDP of the  BH procedure at level $\alpha$ on $\left\lbrace \frac{1}{\hat{\tau}_m}p_j^{r/n}, j\in S_{\hat{\tau}_m} \right\rbrace$ is asymptotically almost surely at most $\alpha$. 
\end{proposition}

\begin{proof}
Consider the following representation of the BH threshold $\hat{t} \equiv \hat{t}(\tau)$ for a given $\tau$ (and a given $\alpha$): 
\begin{equation}
\hat{t}(\tau) = \max\left \lbrace  t \in [0, 1]: \hat{F}_\tau(t) \geq \frac{t}{\alpha}\right \rbrace\label{BH-threshold-ecdf}
\end{equation}
(see, e.\ g., Lemma 5.7 in \cite{Dickhaus-Buch2014}).
 Proposition \ref{propo-BH-fixed-tau-asymptotic-FDR} then yields (under the stated assumptions), that choosing $\hat{t}(\tau)$ as the rejection threshold for the conditional PC $p$-values leads to an FDP which is asymptotically almost surely upper-bounded by $\alpha$ for any fixed $\tau$. In particular, considering $\tau = \tau_\infty$ in \eqref{BH-threshold-ecdf} leads to an FDP which is asymptotically almost surely upper-bounded by $\alpha$, because $\tau_\infty$ is a fixed constant in the interval $(0, 1]$, by assumption. 

Analogously, the BH threshold for a random (selected) value $\hat{\tau}_m$ is given by
\begin{equation}\label{BH-threshold-with-selected-tau}
\hat{t}(\hat{\tau}_m) = \max\left \lbrace  t \in [0, 1]: \hat{F}_{\hat{\tau}_m}(t) \geq \frac{t}{\alpha}\right \rbrace.
\end{equation}
Clearly, this representation implies that $\hat{t}(\hat{\tau}_m)$ depends on the data only via $\hat{F}_{\hat{\tau}_m}$, as soon as $\hat{\tau}_m$ has been chosen. By our assumptions, $\hat{\tau}_m$ converges almost surely to $\tau_\infty$ and $|\hat{F}_{\hat{\tau}_m} - \hat{F}_{\tau_\infty}|$ converges uniformly and almost surely to zero. From these assertions, by arguing similarly to the proof of Lemma A.2 in \cite{finnrot2001}, we conclude that 
$|\hat{t}(\hat{\tau}_m) - \hat{t}(\tau_\infty)|$ converges to zero almost surely as $m$ tends to infinity. However, as argued before, choosing $\hat{t}(\tau_\infty)$ as the rejection threshold for the conditional PC $p$-values leads to an FDP which is asymptotically almost surely upper-bounded by $\alpha$, which yields the assertion of the proposition.
\end{proof}

\begin{remark} $ $
\begin{itemize}
\item[(i)] Notice that the  conditional validity of the PC p-values on which the BH procedure operates is the important assumption apart from \eqref{eq-conv1}--\eqref{eq-conv3} in Propositions \ref{propo-BH-fixed-tau-asymptotic-FDR} and \ref{propo-BH-chosen-tau-asymptotic-FDR}.
In particular, the $p$-values on which the BH procedure operates, say $q_1,\ldots,q_m$, need not correspond to PC null hypotheses, but they need to satisfy that $\left[\frac{1}{\tau}q_i \mid q_i<\tau\right]\geq_{st} Uni[0,1]$ for any $\tau\in (0,1]$ if the null hypothesis corresponding to $q_i$ is true, $i\in \{1,\ldots,m\}$. For example, if such a $q_i$ is a  $p$-value from a true one-sided hypothesis regarding the  parameter of a one-parametric exponential family,  then $q_i\geq_{lr} Uni[0,1]$, which implies conditional validity. This set-up was considered in \cite{Zhao2019}  for control of the family-wise error rate. The results in Propositions \ref{propo-BH-fixed-tau-asymptotic-FDR} and \ref{propo-BH-chosen-tau-asymptotic-FDR} justify applying for such $q_i$'s FDR controlling procedures if an (asymptotic) FDR guarantee is of interest.

\item[(ii)] The assumption that $
\pi_0 G_0(x) + \pi_1 G_1(x)$ has a unique crossing point $x^*(\alpha) \in (0, 1)$ with the "Simes line" $
x / \alpha$ has intensively been discussed by \cite{Chi2007}, \cite{Genovese-Wasserman2002}, and \cite{FiDiRo2007}, among others. It is for instance fulfilled if $
\pi_0 G_0(x) + \pi_1 G_1(x)$ is a strictly concave function fulfilling that 
$$
\lim_{x \downarrow 0} \frac{\pi_0 G_0(x) + \pi_1 G_1(x)}{x} > \frac{1}{\alpha}.
$$
\end{itemize}
\end{remark}

Based on these asymptotic results, we are able to provide an asymptotic FDR control guarantee for our suggested adaptive choice of $\tau$. 
Namely, the FDP of the rejections made with $(\hat \tau, \hat t(\hat \tau))$, as described in \eqref{greedy-tau}, is asymptotically almost surely (a.s.) at most $\alpha$. 

 \begin{corollary}\label{justification-greedy-tau}
Assume that the conditions of Proposition \ref{propo-BH-chosen-tau-asymptotic-FDR} are satisfied. 
Let $\hat \tau$ 
be the greedy choice of $\tau$, as described in \eqref{greedy-tau}. Then, the FDP of the (adaptive) BH procedure at level $\alpha$ on $\left\lbrace\frac{1}{\hat \tau}p_j^{r/n}, j\in S_{\hat \tau} \right\rbrace$ is asymptotically at most $\alpha$. 
\end{corollary}

\begin{proof}
The assertion follows from Proposition \ref{propo-BH-chosen-tau-asymptotic-FDR} since $\hat \tau_m$ depends on the data only via\\ $\{|S_{\tau_i}|/m, \hat F_{\tau_i}, i=1,\ldots, K\}$. Since $|S_{\tau_i}|/m$ and $\hat F_{\tau_i}$  converge almost surely to well defined limiting functions for $i=1,\ldots, K$,  there exists a limiting value $\tau_{\infty} \in \{\tau_1,\ldots,\tau_K \}$  that $\hat \tau_m$ converges to almost surely, and 
$\vert\vert \hat{F}_{\hat{\tau}_m} - \hat{F}_{\tau_\infty} \vert\vert_\infty \to 0$ almost surely as $m \to \infty$.
\end{proof}
 
For practical purposes, the weak dependence conditions stated in \eqref{eq-conv1} can be characterized for several dependence structures which are relevant in genetic applications. To this end, we make use of Theorem 2B on page 420 of \cite{parzen1960} (see also Section 4.4 in \cite{vroni}).
 
 \begin{lemma}
 \label{weak-dep-lemma}
 Let $I_0$ with $|I_0| = m_0$ and $I_1$ with  $|I_1| = m_1$ denote the index sets of true and false PC null hypotheses, respectively. Furthermore, let $\xi_{j} = \mathbf{1}_{[0, t]}(p_{j}^{r/n})$ for $j \in \{1, \ldots, m\}$ and for a fixed value of $t \in [0, 1]$. If 
 \begin{equation}\label{weak-dep-cond}
 \frac{1}{m_\ell} \left| \sum_{j \in I_\ell} \text{Cov}(\xi_{j}, \xi_{\max\{I_\ell\}}) \right| \leq O\left(m_\ell^{-q_\ell}\right)
 \end{equation}
 for both $\ell \in \{0, 1\}$ as $\min\{m_0, m_1\} \to \infty$, for some $q_\ell > 0$ and for all $t \in [0, 1]$, then weak dependence in the sense of \eqref{eq-conv1} can be concluded. In \eqref{weak-dep-cond}, the symbol $O$ refers to Landau's notation.
 \end{lemma}

 \begin{remark}
 The maximum in \eqref{weak-dep-cond} refers to the usual ordering on $\mathbb{N}$. In practice, it is important that the entries in ${\cal P}_{n\times m}$ are arranged such that for each study $i \in \{1, \ldots, n\}$ the resulting ordering of the elementary $p$-values in the $i$-th row of ${\cal P}_{n\times m}$ is meaningful for the application at hand. In the context of genetic association studies, a meaningful ordering can for instance correspond to genetic distance.
 \end{remark}
 
 \begin{corollary}\label{coro-weak-dependence} 
Under the setup of Lemma \ref{weak-dep-lemma}, we may consider the following dependence structures. 
 \begin{itemize}
     \item[(i)] \underline{Block-dependence structure}:  Assume that the $p$-values $(p_{j}^{r/n} : j \in I_\ell)$  are block-dependent for $\ell \in \{0, 1\}$, meaning that the index set $I_\ell$ can be partitioned into blocks (of consecutive indices) such that  $p_{j}^{r/n}$'s belonging to different blocks are stochastically independent (implying that the corresponding $\xi_{j}$'s are   uncorrelated). Then, \eqref{weak-dep-cond} is fulfilled if the maximum block size pertaining to $I_\ell$  is of a smaller (asymptotic) order of magnitude than $m_\ell$; see (4.30) in Theorem 4.16 of \cite{vroni}. Surely, block dependence in each study $i \in \{1, \ldots, n\}$ entails block dependence for the combination $p$-values, if the study-specific blocks are suitably aligned, meaning that all columns of the $n\times m$ matrix of elementary $p$-values exhibit the same or almost the same block-dependence structure.
     \item[(ii)] \underline{Banded covariance structure}: Analogously, assume that the random variables $(\xi_{j}: j \in I_\ell)$ possess a banded covariance structure, meaning that $\text{Cov}(\xi_{j_1}, \xi_{j_2}) = 0$ whenever $j_1, j_2 \in I_\ell$ and $|j_1 - j_2| > k_\ell$ for a banding parameter $k_\ell$, $\ell \in \{0, 1\}$. Then, \eqref{weak-dep-cond} is fulfilled if $k_\ell$ is of a smaller (asymptotic) order of magnitude than $m_\ell$. If a banded covariance structure is present in each study $i \in \{1, \ldots, n\}$ and the study-specific bands are suitably aligned (meaning that all columns of the $n\times m$ matrix of elementary $p$-values exhibit the same or almost the same banded covariance structure), then this entails a banded covariance structure for the $\xi_j$'s. Banded covariance structure holds under M-dependence. 
 \end{itemize}
 \end{corollary}

\section{An application to  Crohn's disease GWAS}\label{sec-Crohn}

We apply \CoFilter in order to identify the single-nucleotide polymorphisms (SNPs) associated with Crohn's disease in at least $r \in \{2,3,4\}$ out of the $n=8$ studies in \cite{franke2010genome}. 
Every study has $m=953{,}241$ autosomal SNPs. Our starting point  is a matrix of $8\times 953{,}241$ two-sided $p$-values for testing that there is no association between SNP $j$ and disease status in study $i$. 

Each $p$-value was obtained from a $2\times 3$ contingency table for the thousands of subjects in each study. Specifics follow. The rows are for the disease status variable (with or without Crohn's disease), the columns are for the number of minor alleles (0, 1, or 2). According to \cite{franke2010genome}, the $p$-value was computed from  a standard 1 degree-of-freedom allele-based test for no association (we did not have access to the $2\times 3$ contingency tables). If there is no association between SNP $j$ and the disease status,  the two-sided $p$-value  is  uniformly distributed (approximately, up to the large sample theory approximation for the null distribution of the test statistic, which is expected to be very good since the studies are large).   

\begin{figure}
 \begin{tabular}{ccc}
 \hspace{-1cm}
 \includegraphics[width=4.5cm,height=4.5cm, page=1]{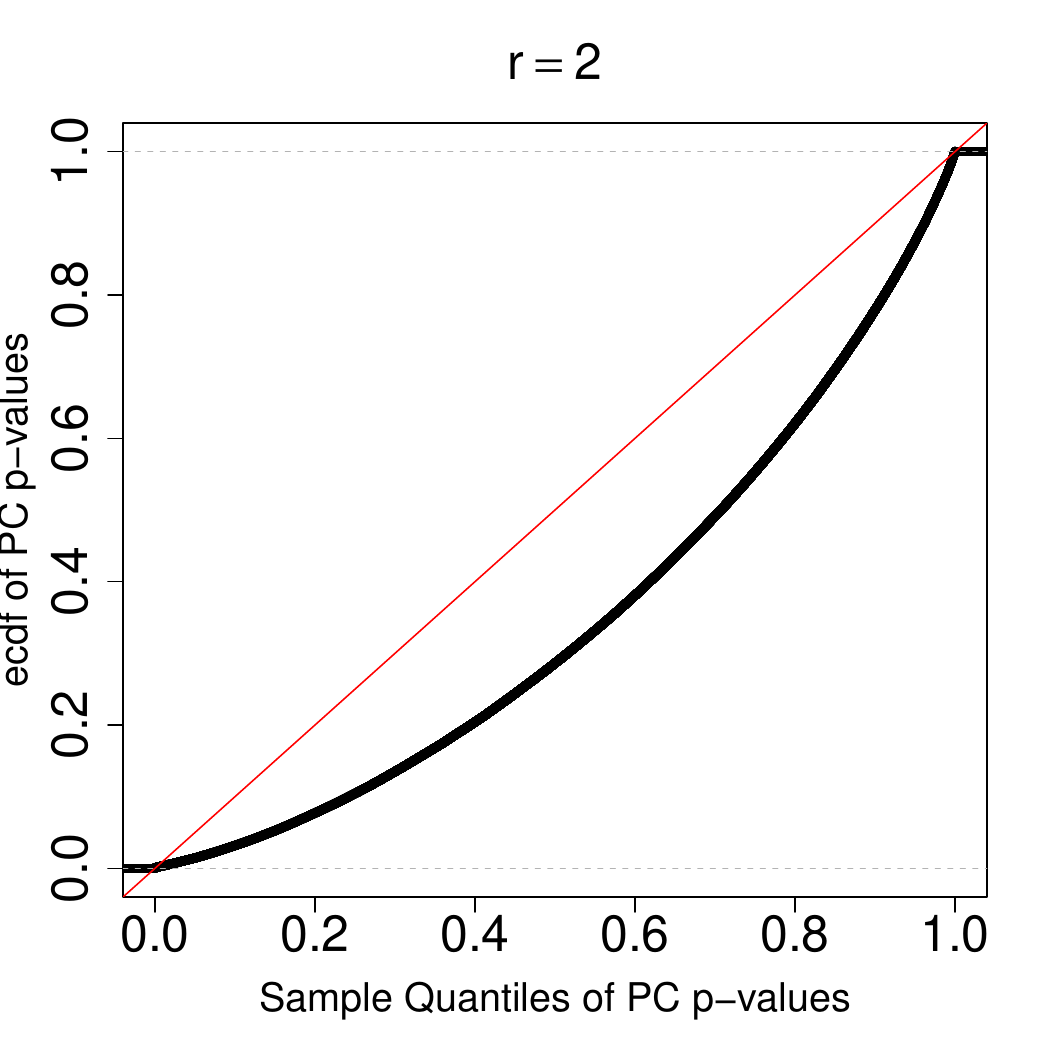} &      \includegraphics[width=4.5cm,height=4.5cm, page=1]{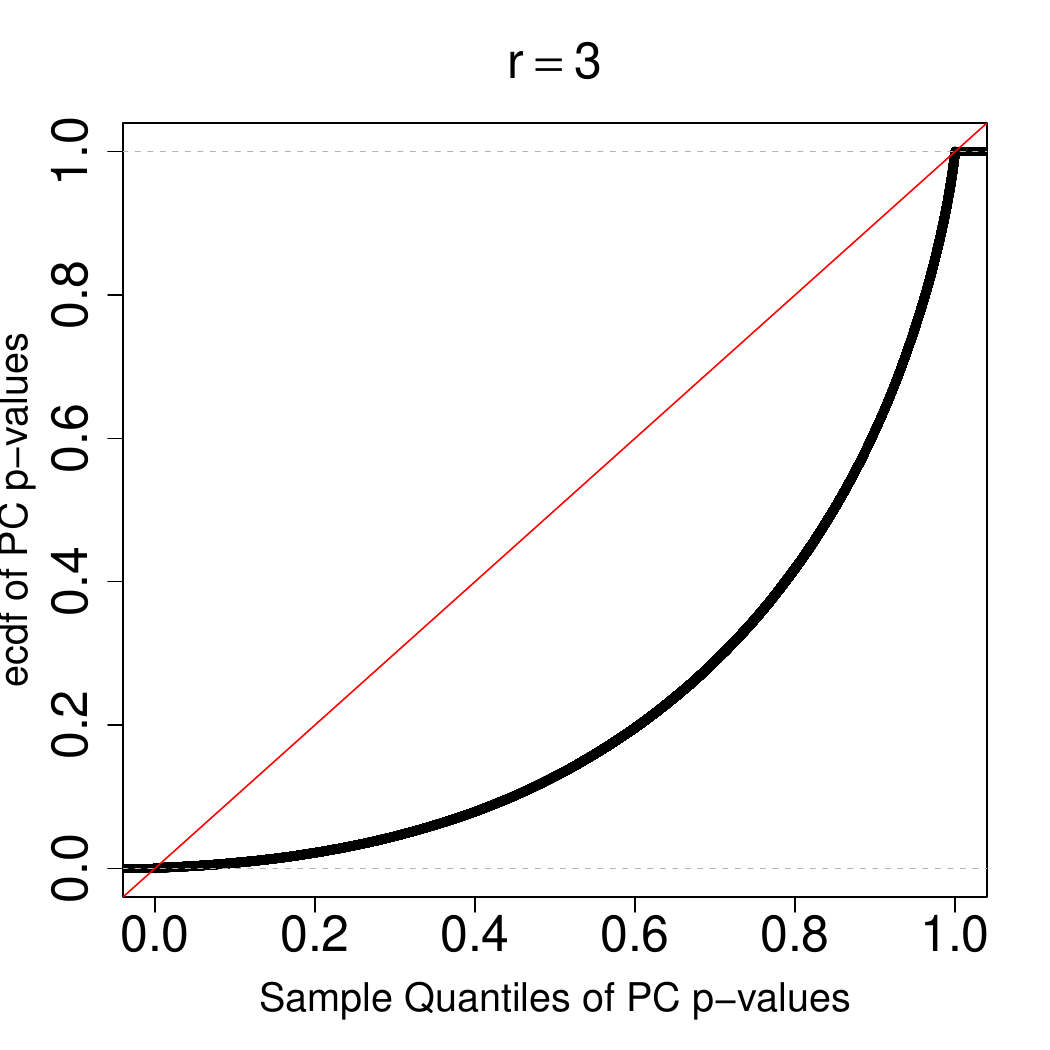} &        
  \includegraphics[width=4.5cm,height=4.5cm, page=1]{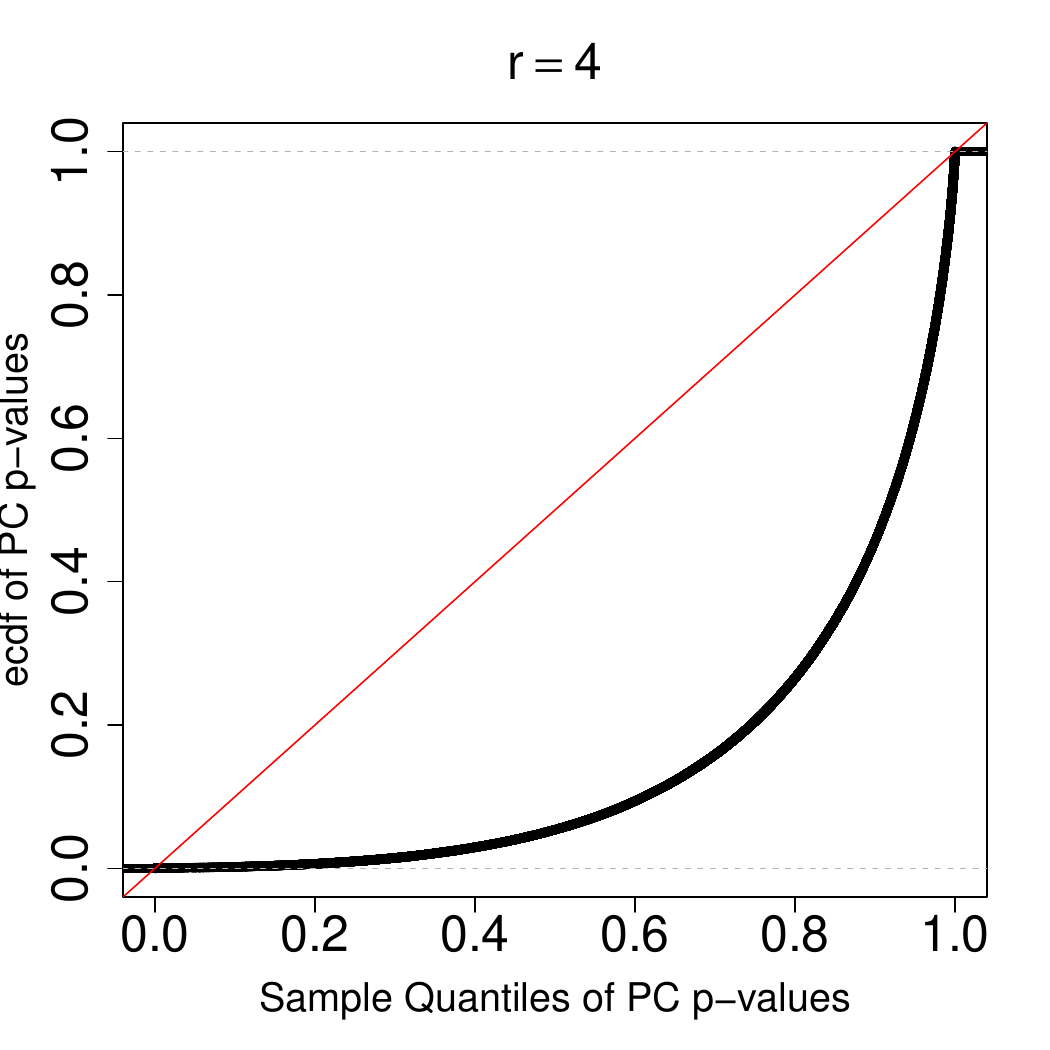}
  \\        
  \hspace{-1cm} \includegraphics[width=4.5cm,height=4.5cm, page=2]{CDecdfgamma2.pdf} &      \includegraphics[width=4.5cm,height=4.5cm, page=2]{CDecdfgamma3.pdf} &        
  \includegraphics[width=4.5cm,height=4.5cm, page=2]{CDecdfgamma4.pdf} 
  \\        
 \hspace{-1cm} \includegraphics[width=4.5cm,height=4.5cm]{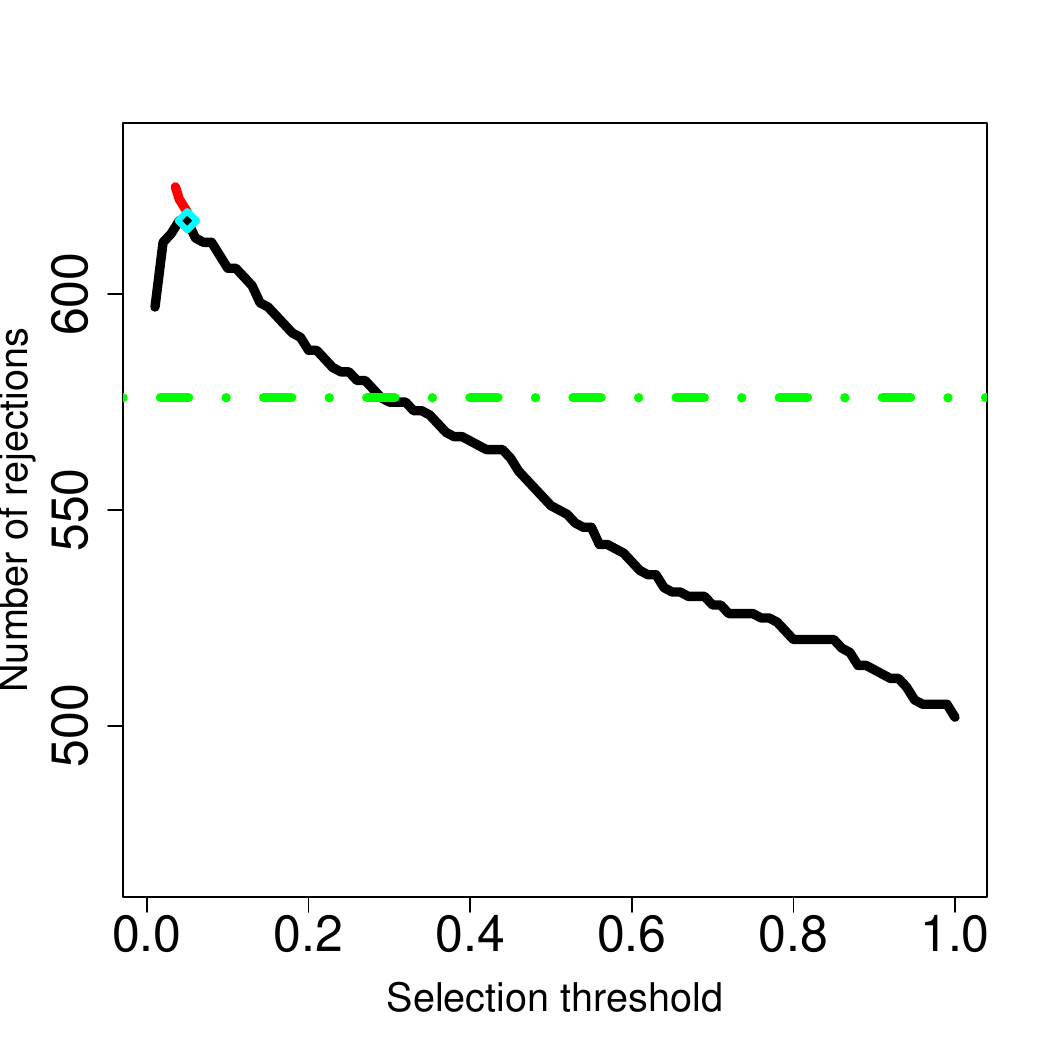} &      \includegraphics[width=4.5cm,height=4.5cm]{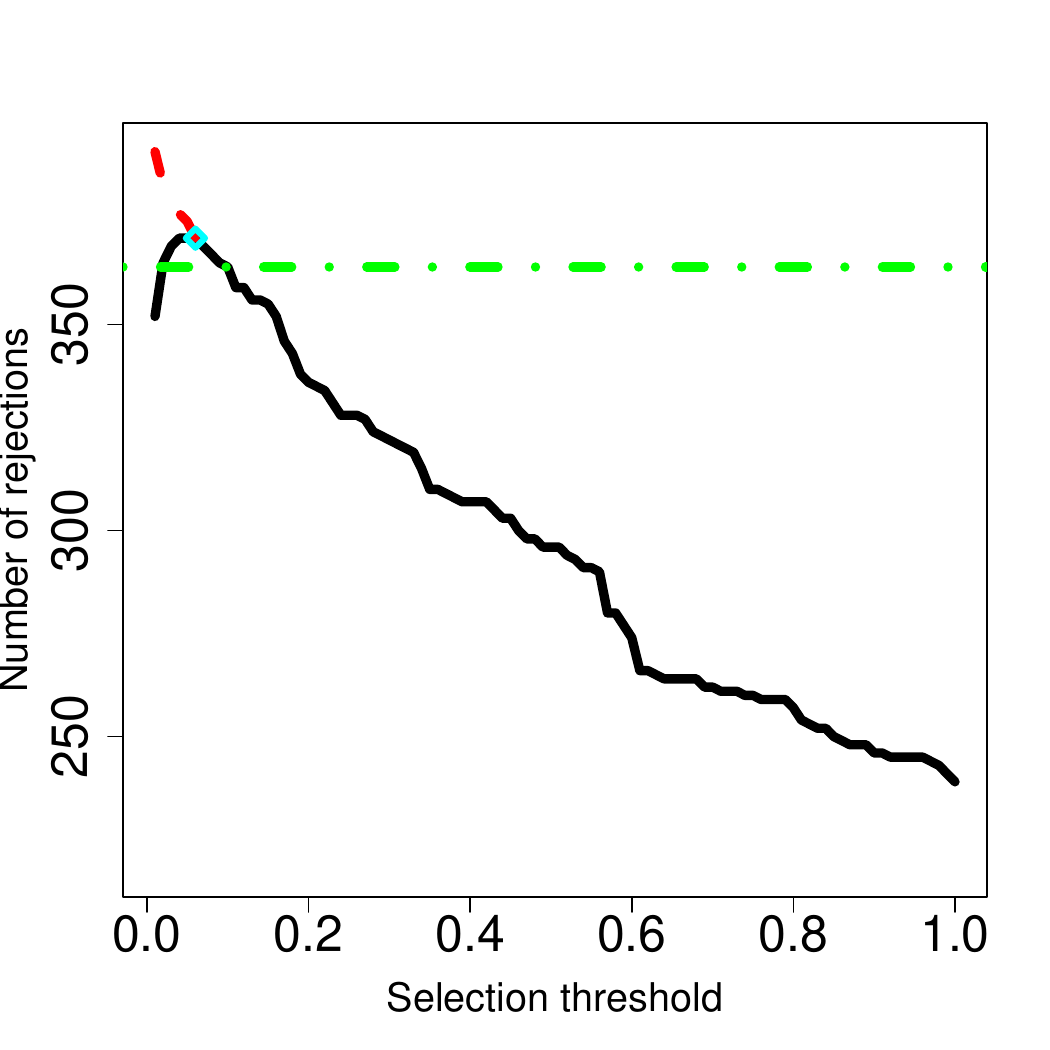} &        \includegraphics[width=4.5cm,height=4.5cm]{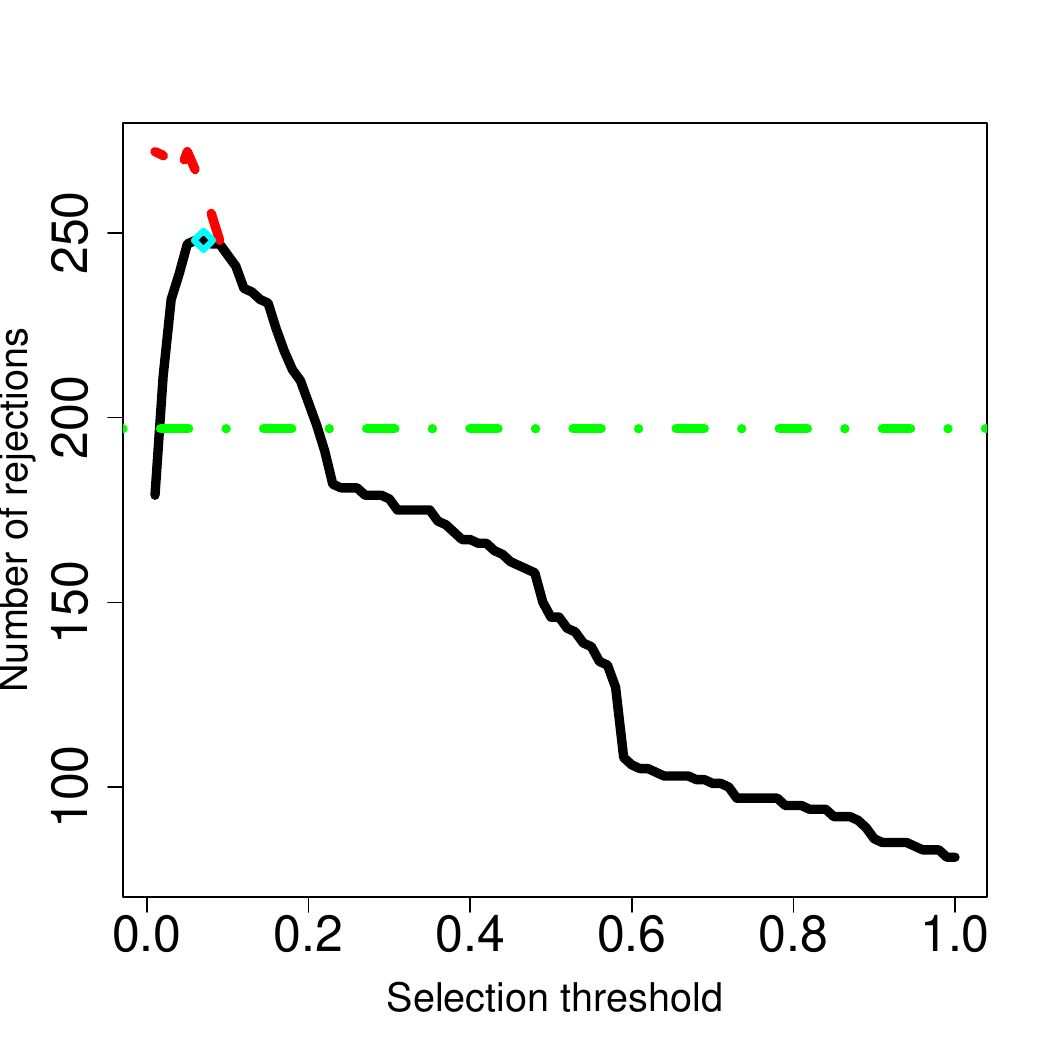} 
  \end{tabular}
 \caption{\label{figCD} For every $r$, the ecdf of the PC $p$-values (top row) and conditional PC $p$-values (middle row), following selection with a pre-specified threshold $\tau = 0.1$. The bottom row shows  the number of rejections at level 0.05, as a function of the selection threshold,   using the conditional $p$-values for BH (solid black curve) and adaptive BH (dashed red curve, displayed only for $\tau$ values for which the estimated fraction of null PC hypotheses below one), as well as for adaFilter (dash-dotted green line).  The number of rejections using the greedy threshold \eqref{greedy-tau} is the cyan diamond point. The values at selection threshold `1.0` correspond to no selection, i.e., the BH  on the (unconditional) PC $p$-values. 
}
 
\end{figure}

The PC $p$-values are computed using the Fisher combination method on the two-sided $p$-values. Their empirical cumulative distribution function (ecdf), displayed in the top row of Figure \ref{figCD}, suggests that  the majority of the PC $p$-values have a distribution that is stochastically greater than uniform, as they are below the 45-degree diagonal line (and the deviation from uniformity increases with $r$).  Near the origin, we expect to find the PC $p$-values from false PC null hypotheses that have a distribution that is stochastically smaller than $Uni[0,1]$. The ecdf of the conditional PC $p$-values, in the middle row of Figure \ref{figCD}, shows that the distribution is concave near the origin, suggesting that there is a non-negligible proportion of false PC null hypotheses in the data. The middle row figures can be considered a zoom-in of the respective top row figures on $[0,\tau]\times[0,\hat{F}(\tau)]$, where $\hat{F}$ denotes the ecdf.

For FDR control at the 0.05 level, we apply \texttt{CoFilter} with the BH procedure or the adaptive BH procedure on the conditional $p$-values in step (iii). The threshold for selection is either a pre-specified fixed value or adaptively chosen from the grid of values $0<0.01<0.02<\ldots<1$ as defined in \eqref{greedy-tau}. 

For comparison, we also apply on the (unconditional) PC $p$-values the 0.05 level BH and adaptive BH. They are referred to as the unconditional methods. These methods correspond to applying \texttt{CoFilter} with $\tau=1$ and in step (iii) the 0.05 level BH or adaptive BH. In addition, we apply the competitor that is most similar to our approach, AdaFilter \citep{Wang20}. 
Their method is based on the Bonferroni combination method for testing PC $p$-values, using the following clever approach. Let $p_{j(1)}\leq \ldots \leq p_{j(n)}$ be the order statistics of the $n=8$ $p$-values for SNP $j$. Then $(n-r+1)p_{j(r-1)}$ and $(n-r+1)p_{j(r)}$ are, respectively, valid $p$-values for testing $H_j^{(r-1)/n}$ and $H_j^{r/n}$. They  use $F_j=(n-r+1)p_{j(r-1)}$ 
 as the filtering statistic  and  $S_j=(n-r+1)p_{j(r)}$ as the selection statistic
 for feature $j\in\{1, \ldots, m\}.$ Their selection threshold is
$$\gamma_0^{BH}=\sup\Big\{\gamma\in[0, \alpha]\,\Big|\,\frac{\gamma\sum_{i=1}^m I(F_i<\gamma)}{\max(\sum_{i=1}^mI(S_i<\gamma), 1)}\leq \alpha\Big\},$$
so they reject $H_j^{r/n}$ if $S_j<\gamma_0^{BH},$ for $j\in\{1, \ldots, m\}.$

The bottom row of Figure \ref{figCD} shows the number of rejections by each method for each $r\in\{2,3,4\}$. Our novel approach makes more discoveries than the unconditional methods that apply the BH or adaptive BH procedure on the (unconditional) PC $p$-values. Table \ref{tab4} shows the number of selected PC hypotheses and the estimated fraction of null PC hypotheses among the selected when the selection threshold is pre-specified at $\tau=0.1$, as well as when it is adaptively chosen, for each $r$. 
 Since most $p$-values have a null distribution that is stochastically much larger than uniform, selection tends to eliminate many more PC null hypotheses than the number expected  had all the PC null $p$-values been uniformly distributed. 
 Since the number selected divided by the threshold for selection, $|S_{\tau}|/\tau$, is  much smaller than $m=953,241$ in every row of  Table \ref{tab4}, we expect  the threshold for discovery for each selected hypothesis to be lower using the conditional approach. To see this, consider first the effect of filtering on the simpler Bonferroni procedure in step (iii) of \texttt{CoFilter}. Each selected hypothesis would have  been rejected if the conditional PC $p$-value is at most $\alpha/|S_{\tau}|$ or the (unconditional) PC $p$-value is at most $\alpha\times \tau/|S_{\tau}|$. The standard approach is to reject if the (unconditional) PC $p$-value is at most $\alpha/m$. Since  $\alpha\times \tau/|S_{\tau}|>\alpha/m$, the conditional approach will result in at least as many rejections as the standard approach.  This reasoning carries over also to BH instead of Bonferroni: if all the false nulls (with enough power for detection) are selected, \texttt{CoFilter} has more power than the unconditional methods.

The bottom row of Figure \ref{figCD} also shows that when applying \texttt{CoFilter} with $\hat \tau$ as described in \eqref{greedy-tau}, we  make more discoveries than adaFilter. This is also the case when using \texttt{CoFilter} for fixed  $\tau$ in some range. This application is particularly suited for  \texttt{CoFilter}, since the filtering step is extremely effective and thus compensates for the necessary adjustment of the PC $p$-values after filtering. adaFilter is more sensitive than \texttt{CoFilter} to the ratio of the number hypotheses that can be rejected when testing  the  $H^{r/n}_j$'s out of the number of hypotheses that can be rejected when testing the  $H^{(r-1)/n}_j$'s, since it uses a less efficient test statistic (based on the Boferroni combining function, which  unlike the Fisher combining method,  does not pool the information across studies by summation): the larger this ratio is, the greater the power of adaFilter.  In Section \ref{sec-simulations} we show that which of  \texttt{CoFilter} and adaFilter is the more powerful procedure  depends on  this ratio.

Applying step (iii) of  \texttt{CoFilter} the adaptive BH procedure can lead to more rejections  than when applying  BH, only if the  estimated fraction of  true null  PC hypotheses among the selected  (i.e.,  with PC $p$-values at most the selection threshold $\tau$) is below one. This was the case  only when the selection threshold was small enough (the dashed red curve in Figure \ref{figCD}). Few additional rejections  were then made with adaptive BH (compared to BH) in step (iii) of \texttt{CoFilter}.

\begin{table}
\caption{\label{tab4} In the Crohn's disease data, for each $r$:  the threshold for selection (column 3), number selected (column 4), and 
estimated fraction of null PC hypotheses among the selected (column 5), for the following  methods for threshold selection: pre-specified at 0.1,  and adaptively selected (as detailed in Section \ref{sec2:multiple}).}
\begin{tabular}{clccc}
  \hline
 $r$  & threshold selection method &$\tau$ &$|S_\tau|$&  estimated null fraction in $\mathcal S_{\tau}$      \\ 
  \hline
2& pre-specified at $\tau=0.1$ & 0.10 & 30385 & 1.09 \\ 
  & adaptive & 0.04 & 11012 & 0.952 \\ 

\hline  
  3&pre-specified at $\tau=0.1$ & 0.10 & 7328 & 1.13 \\ 
  & adaptive & 0.05 & 3188 & 0.943 \\
 \hline 
 4&pre-specified at $\tau=0.1$  & 0.10 & 1973 & 1.05 \\ 
   & adaptive& 0.06 & 1094 & 0.737 \\ 
 \hline 
\end{tabular}
\end{table}

\section{Simulations}\label{sec-simulations}

In this section, we aim to compare  the FDR and power of \texttt{CoFilter},  the unconditional methods, and adaFilter \citep{Wang20}. For \texttt{Cofilter}, we shall  examine how the choice of a fixed selection threshold $\tau$ matters, as well as assess the performance of the greedy choice $\hat \tau$ described in \eqref{greedy-tau}.

We consider the following data generation setting, which is partly inspired by the Crohn's disease  example considered in Section \ref{sec-Crohn}, where we analyse  $n=8$ GWAS studies, and it is estimated that about 90\% of the SNPs have no  association with the disease in all the studies. 
We set the number of PC hypotheses based on $n=8$ independent studies to $m=50000$, so we examine $\{H_j^{r/8}, j=1, \ldots, 50000\}$, for $r\in \{2,3,4 \}$. 
 The fraction of true global null hypotheses (i.e., for which all $n$ elementary hypotheses are true) is 0.9.  The fraction of false null PC hypotheses is $\pi_1 \in \{0.001, 0.01\}$, where each false null PC hypothesis has equal probability of being any of the possible false null PC configurations (where a false null PC configuration has $\geq r$ false elementary null hypotheses). The remaining fraction of true null PC hypotheses is $1-0.9-\pi_1$. Each of the remaining true null PC hypotheses  has equal probability of being any of the possible true null PC configurations except the global null configuration (i.e., the configuration has between 1 and $r-1$ false elementary null hypotheses). This type of data generation was considered in \cite{Wang20} and made available in their function  
{\it GenPMat()} in their  R package implementing AdaFilter  available at \url{https://github.com/jingshuw/adaFilter}. We use their package both for generating data from the Gaussian shift model described next and for the adaFilter analysis.

Within each study the test statistics are Gaussian, with a block correlation structure.  The covariance within each block is symmetric   with off diagonal entry $\rho=0.9$ and diagonal entry one and the  block size is 10 (block sizes up to 200  and other correlation strengths were also examined, but since they produced very similar results they are not shown). 
For elementary null hypotheses that are false, the  means are sampled independently from $\{3,4\}$. For elementary null hypotheses that are true, the mean is zero.

We apply \texttt{CoFilter} for a selection threshold $\tau\in (0,1)$. 
 In step (i),  the PC $p$-values are computed using the Fisher combination method.
 In step (ii), $\tau$ is either fixed for all hypotheses at a pre-specified value, or chosen adaptively as described in \eqref{greedy-tau}. 
In step (iii) of \texttt{CoFilter}, the following two multiple testing procedures are applied to the conditional $p$-values for FDR control at the 0.05 level: the BH procedure at level 0.05; the adaptive BH procedure  which applies the BH procedure at level $0.05/\hat \pi_0(S_{\tau})$, where $\hat \pi_0(S_{\tau})$ is defined in \eqref{eq-pi0hat}.   

\begin{figure}
 \begin{tabular}{ccc}
  \hspace{-1cm}\includegraphics[width=4.5cm,height=4.5cm]{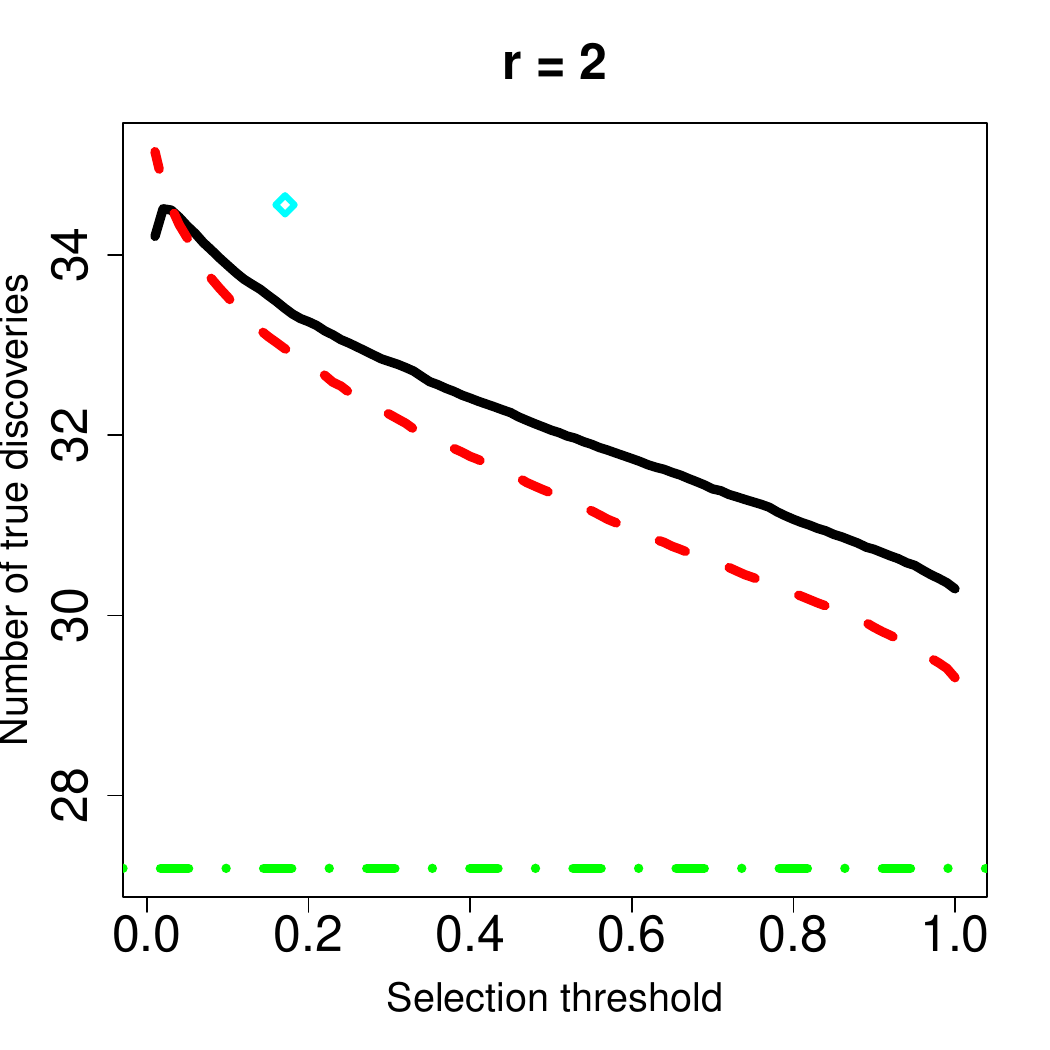} &      \includegraphics[width=4.5cm,height=4.5cm]{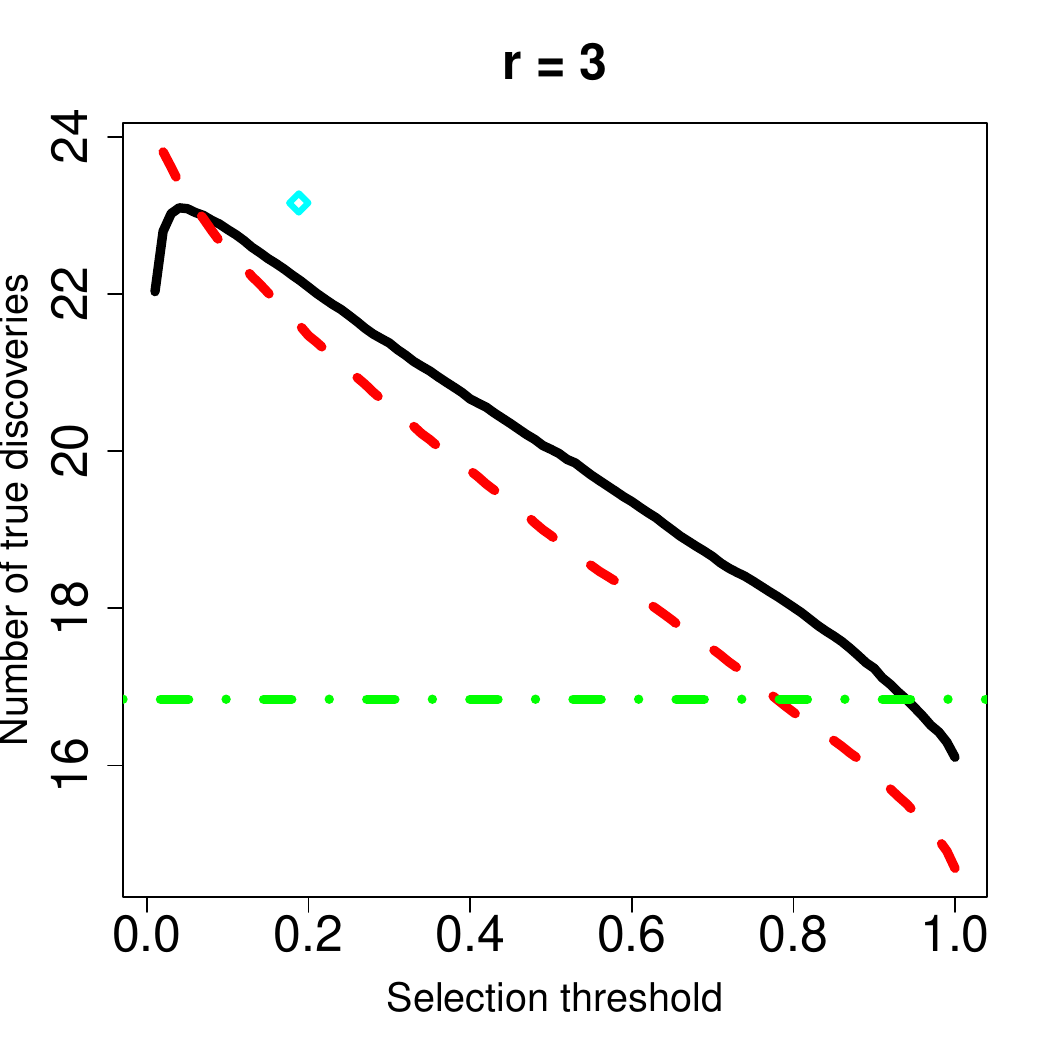}        & \includegraphics[width=4.5cm,height=4.5cm]{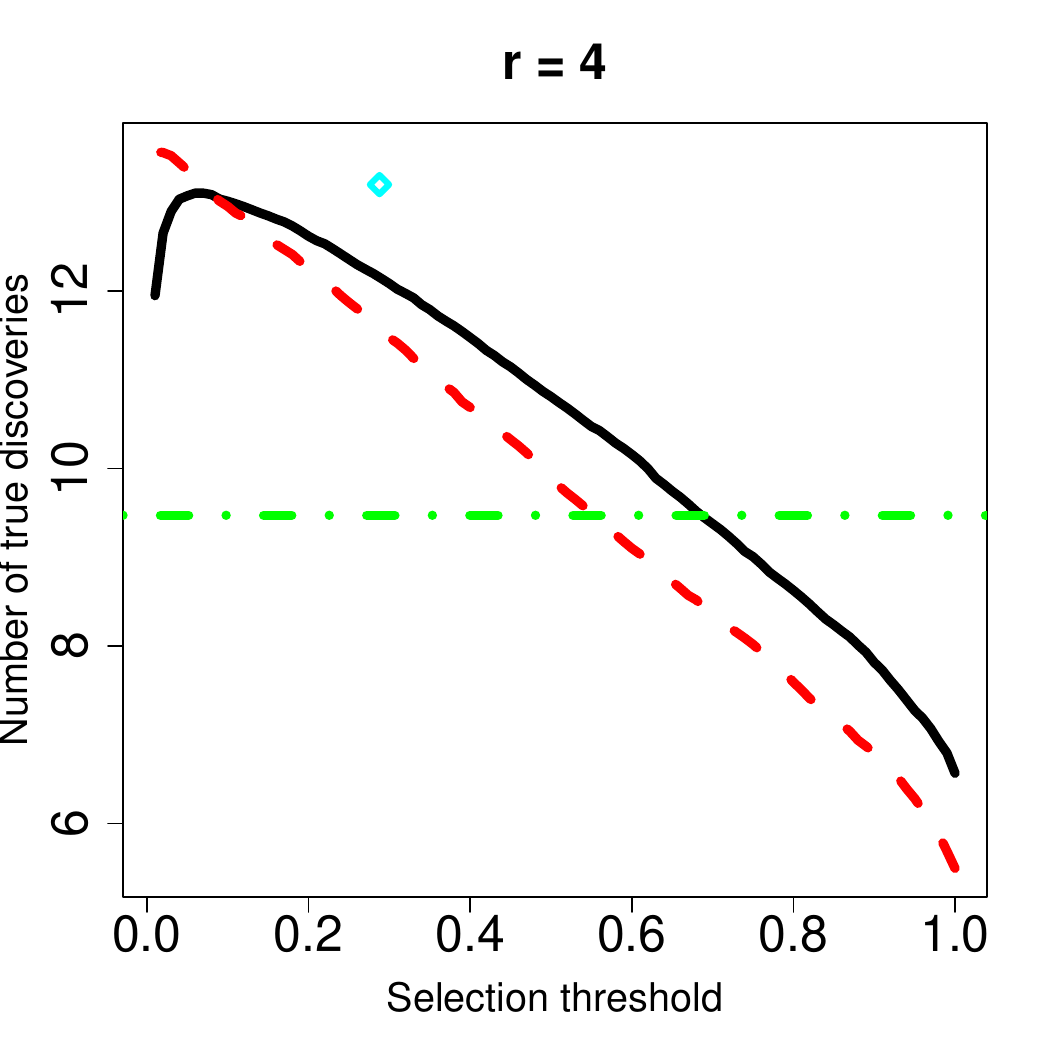} 
  \\ 
 \hspace{-1cm} \includegraphics[width=4.5cm,height=4.5cm]{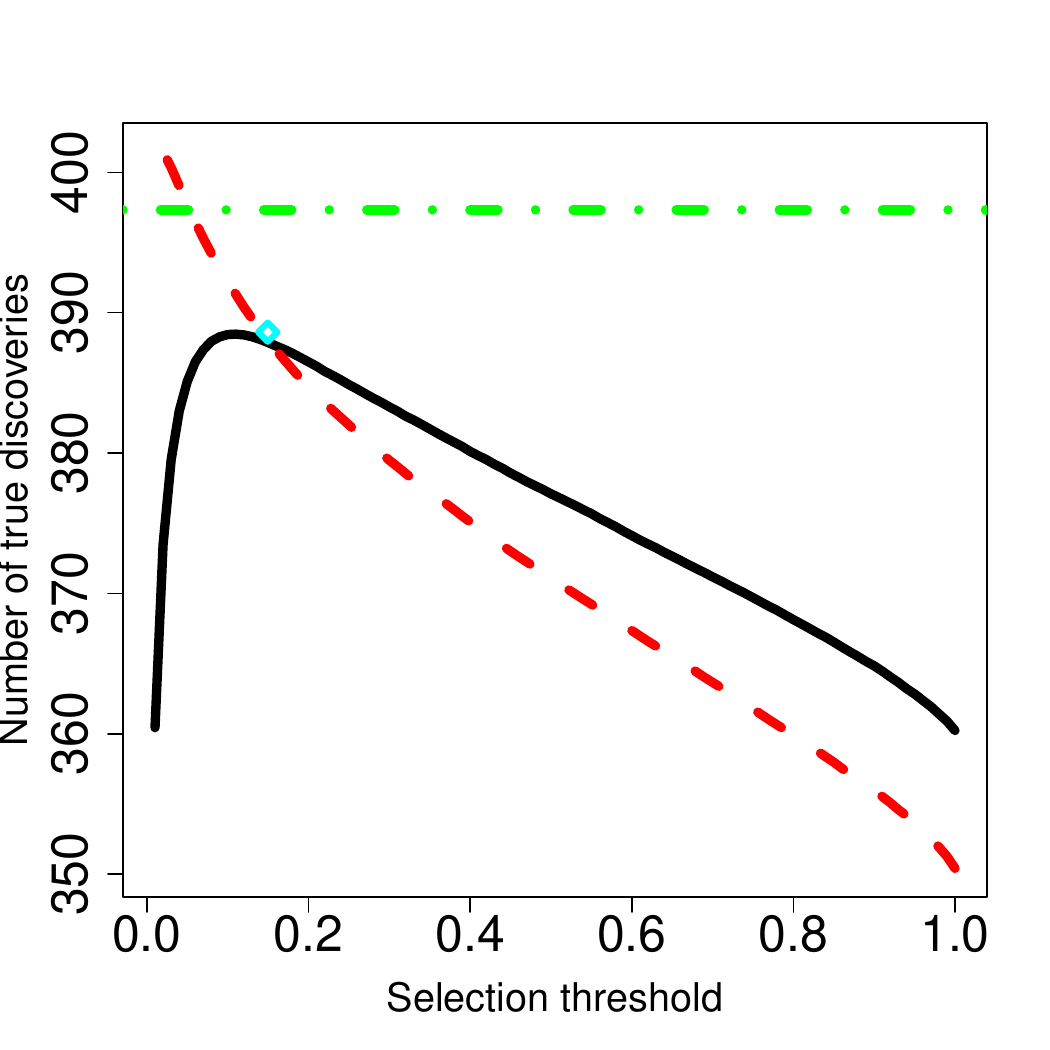} &      \includegraphics[width=4.5cm,height=4.5cm]{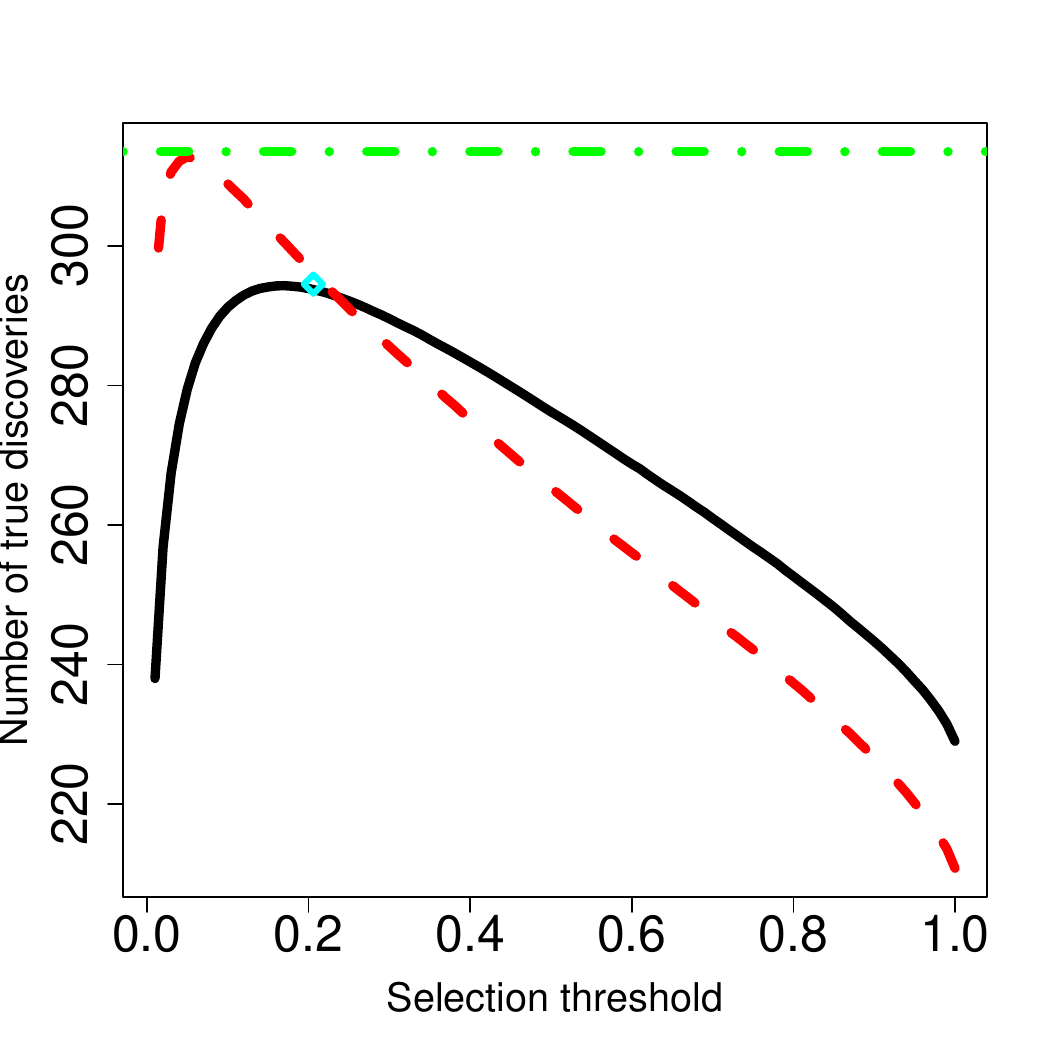}  & \includegraphics[width=4.5cm,height=4.5cm]{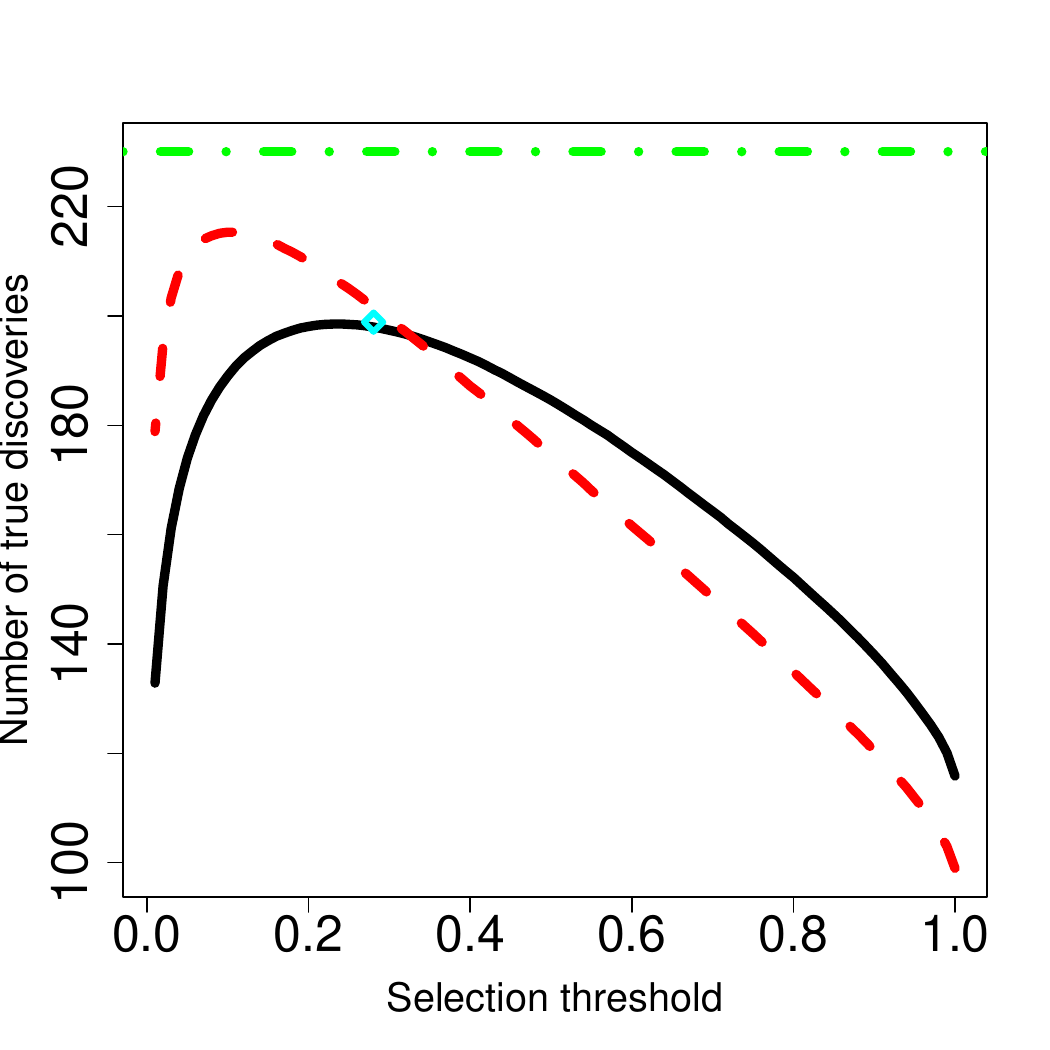} 
 \\
  \hspace{-1cm}\includegraphics[width=4.5cm,height=4.5cm]{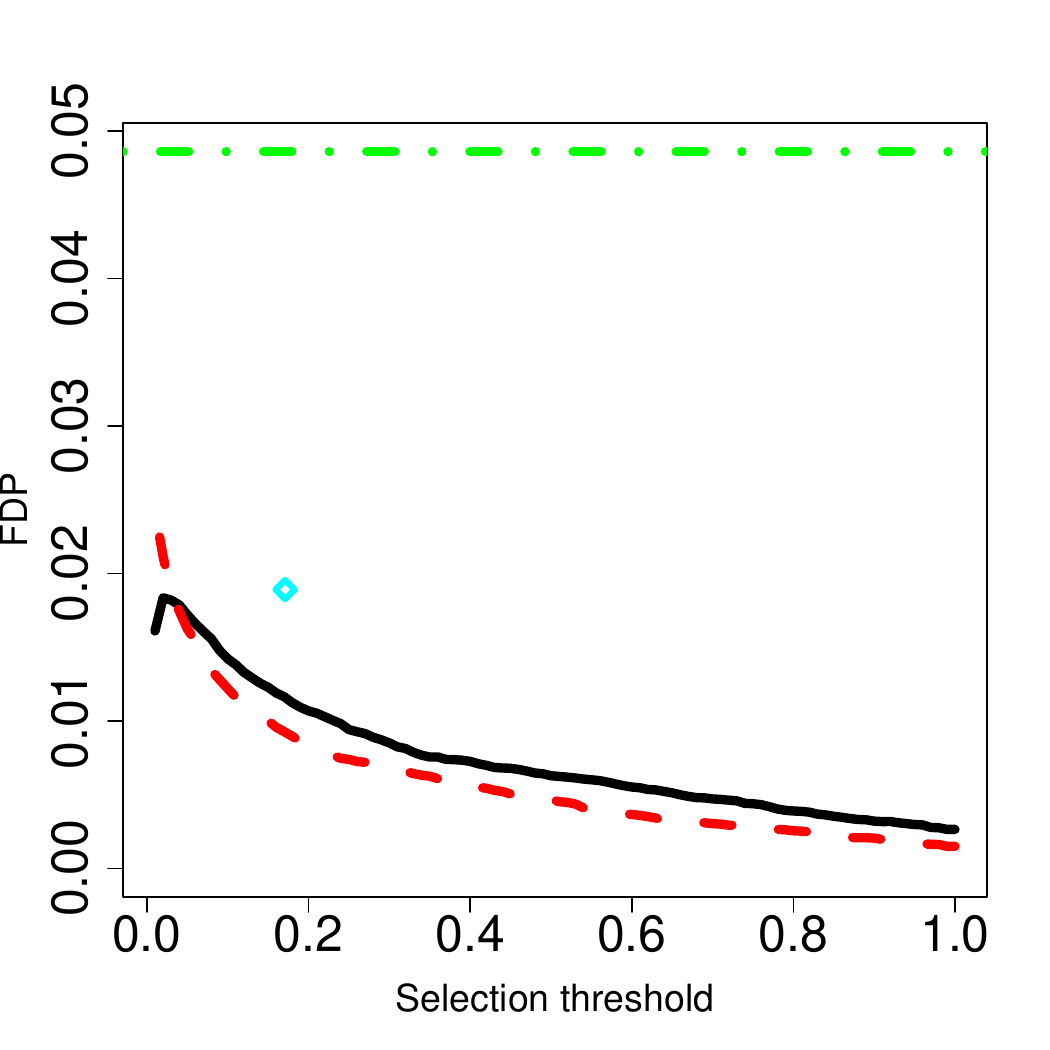} &      \includegraphics[width=4.5cm,height=4.5cm]{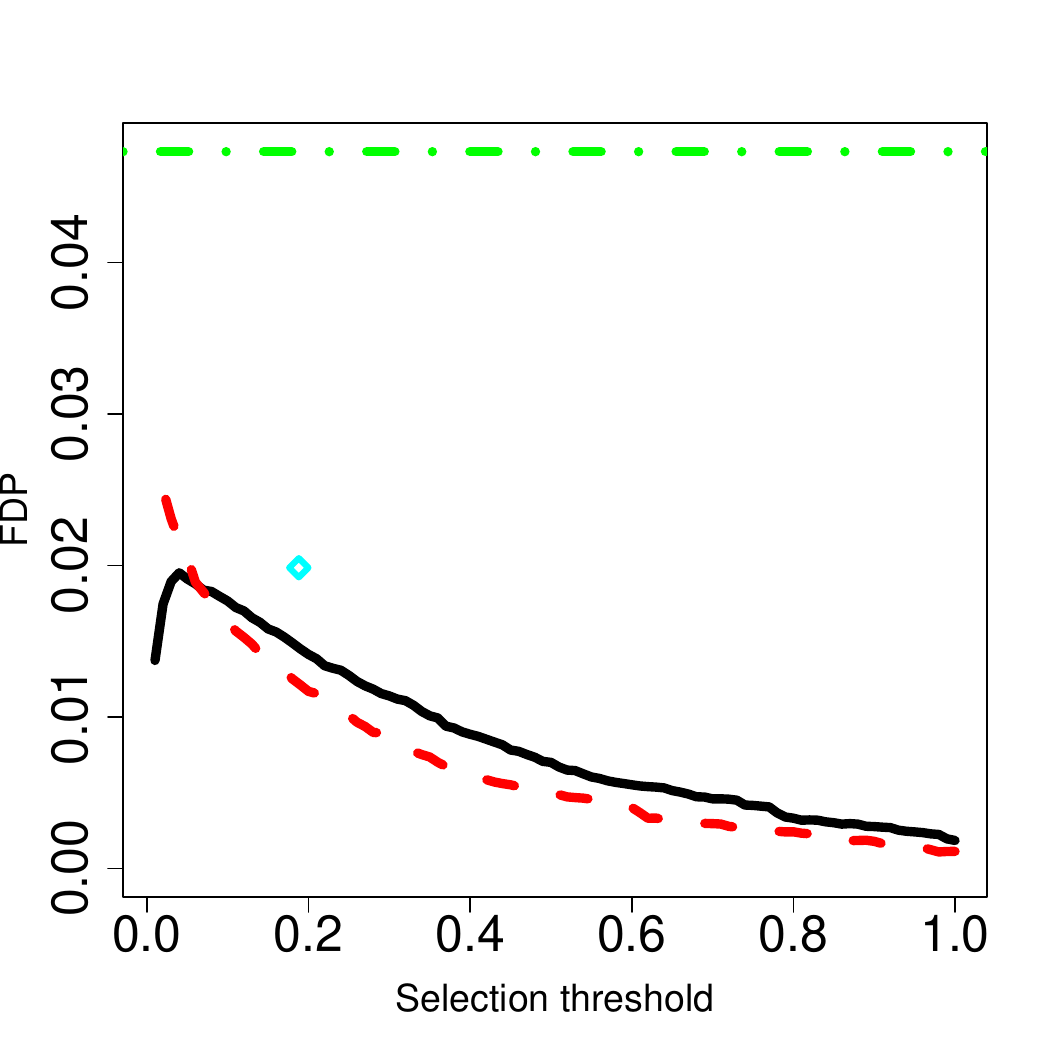}        & \includegraphics[width=4.5cm,height=4.5cm]{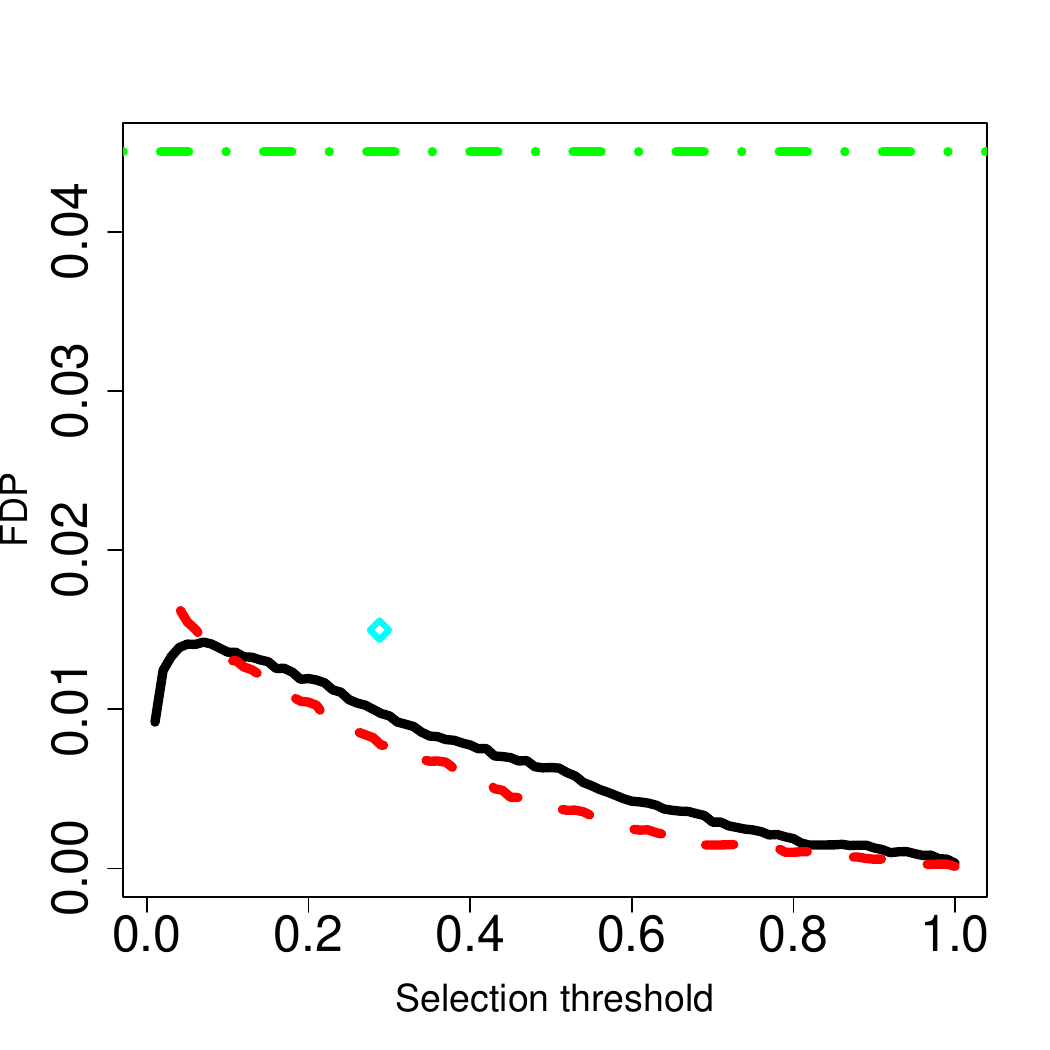}
  \\ 
 \hspace{-1cm} \includegraphics[width=4.5cm,height=4.5cm]{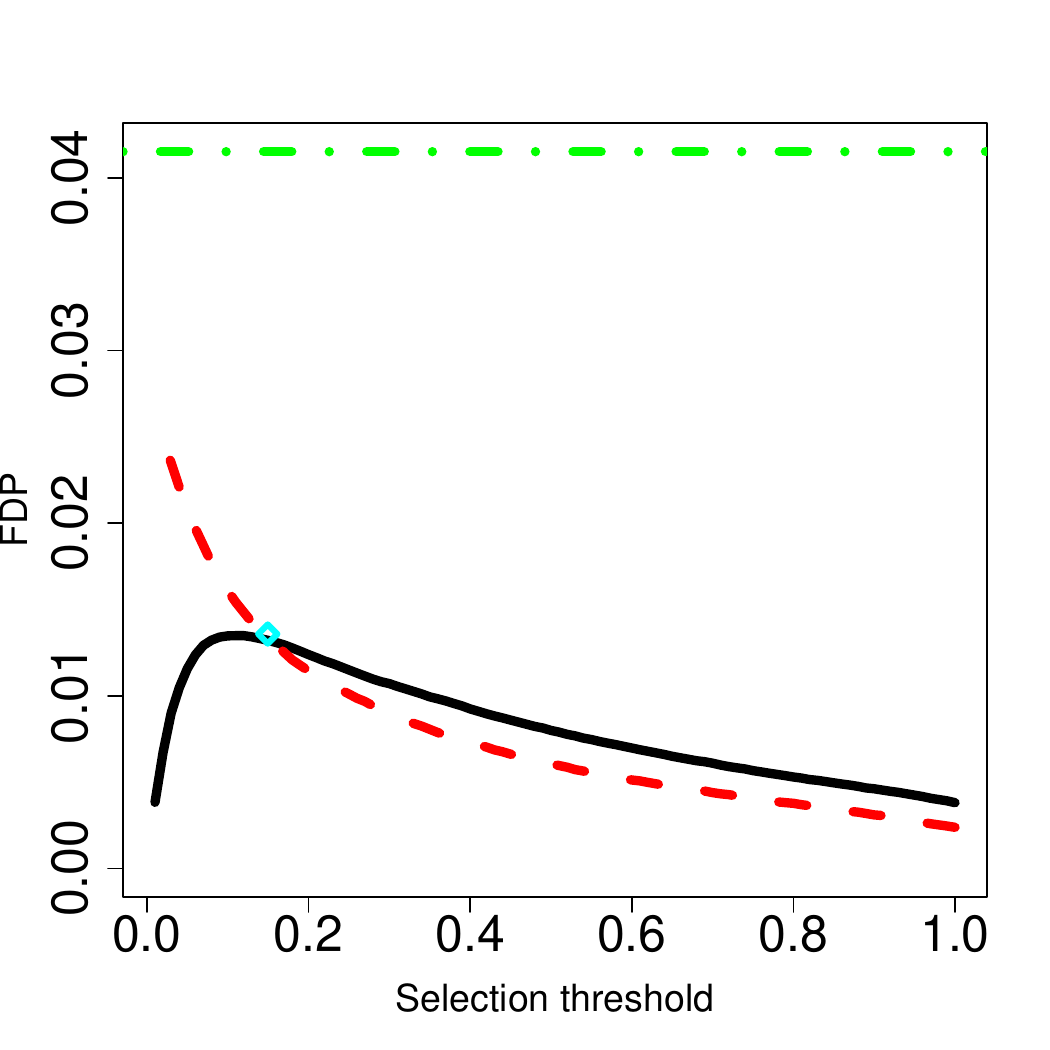} &      \includegraphics[width=4.5cm,height=4.5cm]{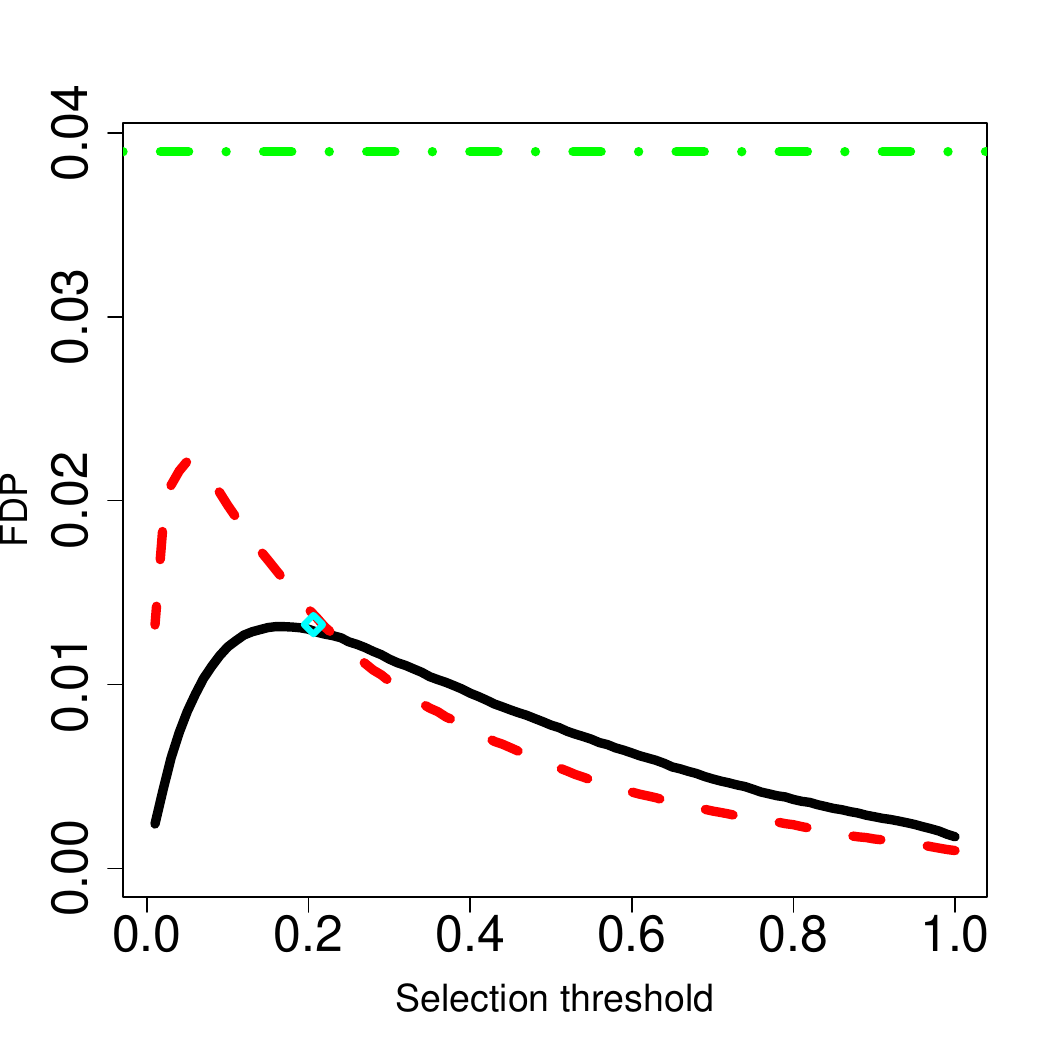}  & \includegraphics[width=4.5cm,height=4.5cm]{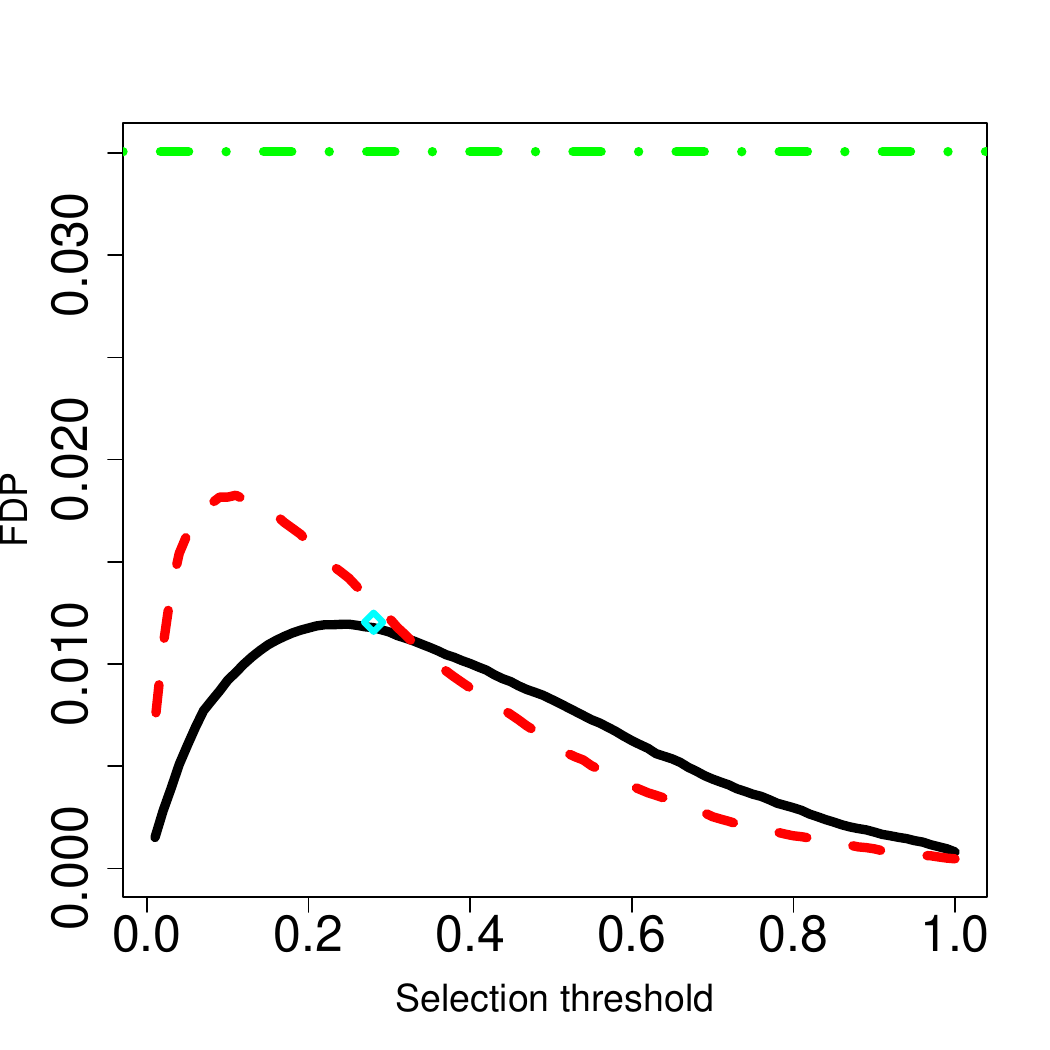} 
  \end{tabular}
 \caption{
  In the symmetric block dependent setting, for each $r\in \{2,3,4\}$ (columns)
   using the conditional $p$-values for BH (solid black) and adaptive BH (dashed red), and adaFilter (dash-dotted green), at level 0.05:
  the average number of true discoveries (rows 1--2) or FDP (rows 3--4) versus the selection threshold.  Cyan diamond is the average value for BH using the greedy choice of $\tau$, at the average greedy $\tau$ value. 
 The fraction of false null PC hypotheses, $\pi_1$, is 0.001 in rows 1 and 3 and  0.01 in rows 2 and 4.  
 Based on 1000 repetitions.
}
\label{fig-sim}
\end{figure}

Figure \ref{fig-sim} shows the average number of true discoveries (our notion of power), and average FDP, using the different methods. Compared with the unconditional methods (which corresponds to $\tau=1$) of applying BH on the PC $p$-values, we see that the power is greater with the novel approach, for every $r$. 
 From the last two rows in Figure \ref{fig-sim}  it is clear that the conditional and unconditional methods are below the nominal 0.05 level, and that the unconditional methods are very conservative (i.e., with low FDR level). This conservatism is due to the fact that most PC $p$-values have a null distribution that is stochastically much larger than uniform. The conditional $p$-values have a null distribution that is closer to uniform, but still conservative.

Compared with adaFilter, we see  that the power of \texttt{CoFilter} is greater  when $\pi_1 = 0.001$ but not when $\pi_1=0.01$.
The setting with $\pi_1=0.01$ is more favorable to adaFilter (over $\pi_1 = 0.001$), since the ratio of  the number of $H^{r/n}_j$'s that are false to the number of $H^{(r-1)/n}_j$'s that are false is larger (due to the fact that the number of true global null hypotheses is unchanged as $\pi_1$ increases).    
In Figure \ref{fig-sim}, bottom two rows, we see that  adaFilter controls the FDR but its FDR is much larger  than the other procedures in all settings considered.

\begin{figure}
 \begin{tabular}{cc}
  \includegraphics[width=6cm,height=6cm, page=1]{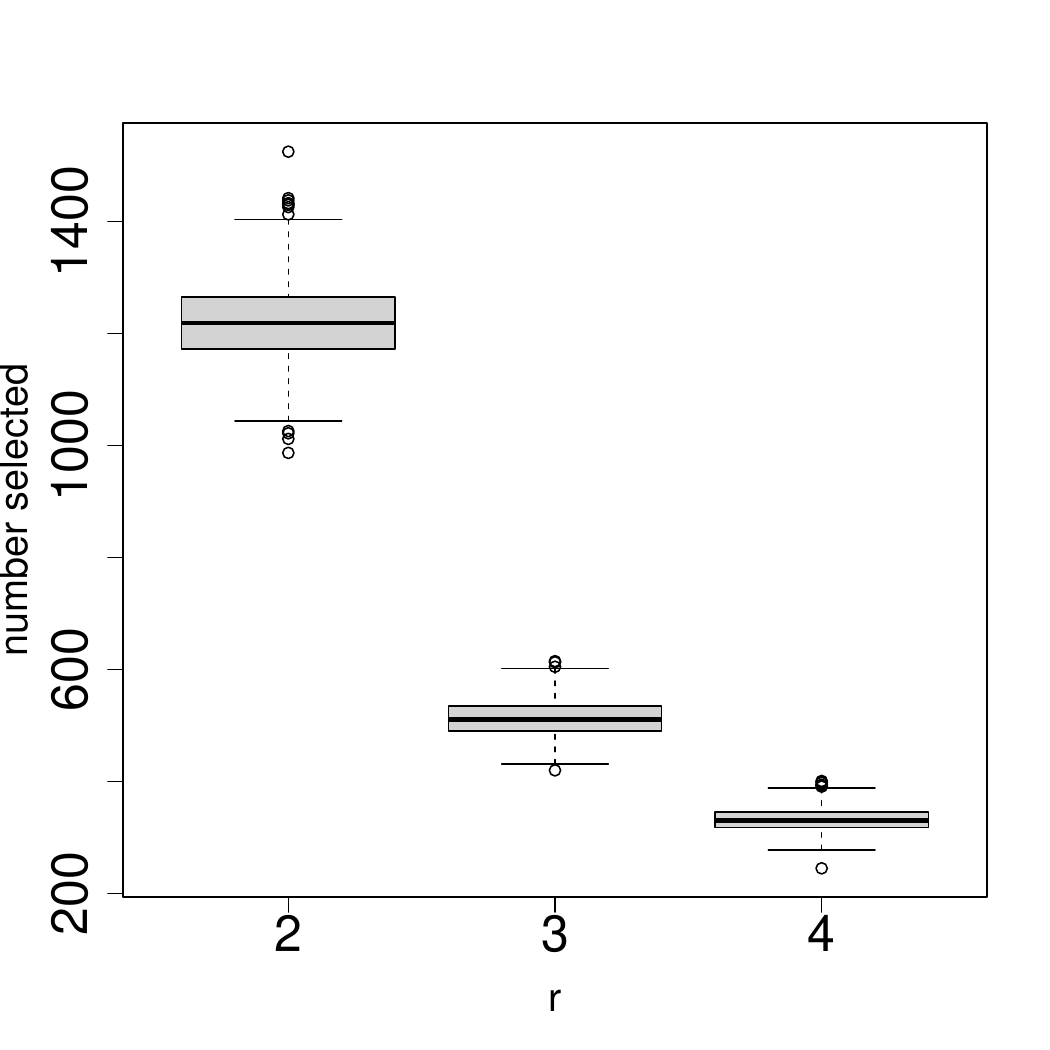} &      \includegraphics[width=6cm,height=6cm, page=1]{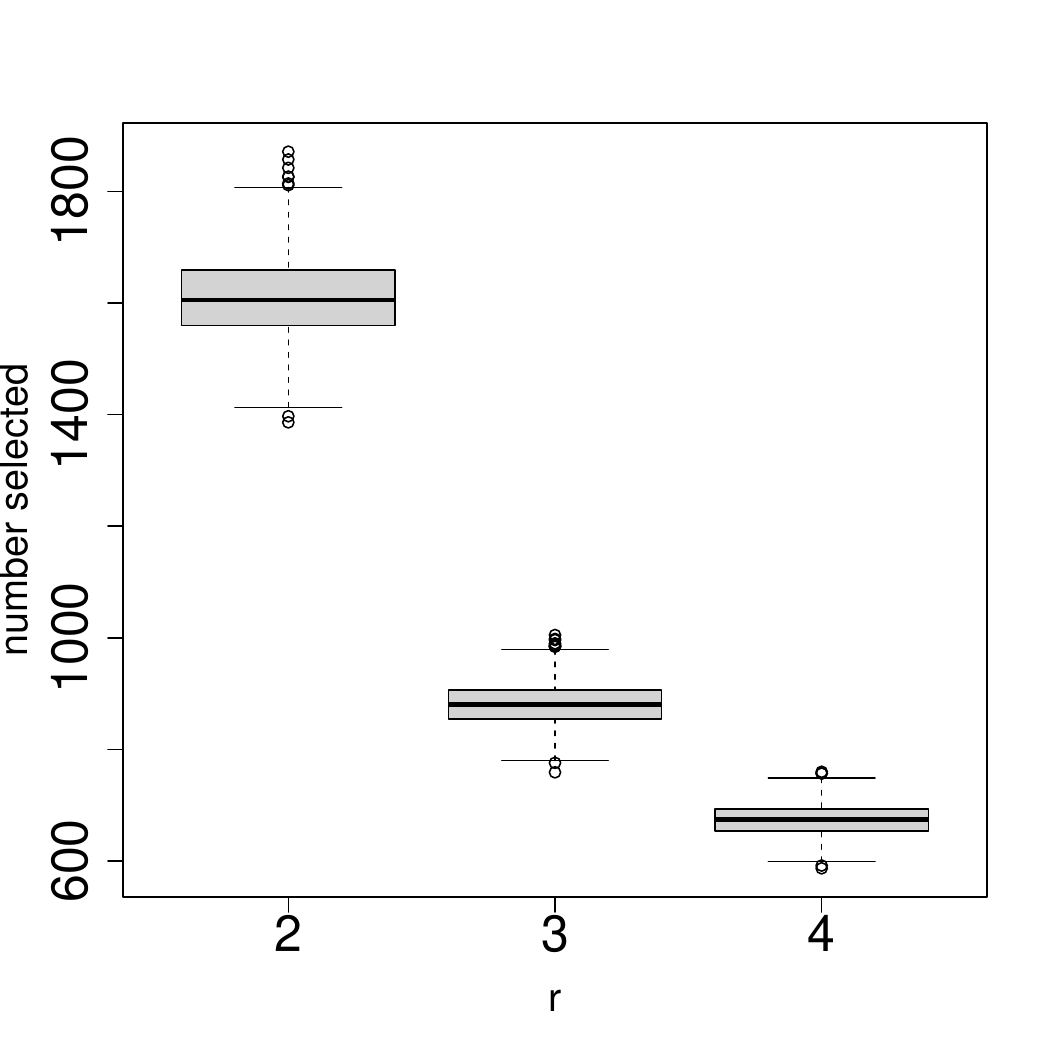} 
  \end{tabular}
 \caption{\label{fignumberselected} The distribution of $|S_{\tau}|$, for every $r$, in the simulation with $\pi_1=0.001$ (left panel) and $\pi_1 = 0.01$ (right panel), for  $\tau=0.1$.  Based on 1000 repetitions.} 
\end{figure}

Since the fraction of true PC hypotheses is close to one, the adaptive BH procedure on the PC $p$-values does not have a power advantage over BH. We see  that when $\pi_1 = 0.001$,   the adaptive approach has a power advantage over BH on the conditional PC $p$-values only when the selection threshold is small, but when $\pi_1 = 0.01$,   the adaptive approach makes more discoveries on average over a wider range of selection thresholds. 
 This is so because the advantage of adaptivity increases as $|S_{\tau}|$  contains a smaller fraction of nulls. For example, for $r=2$, Figure \ref{fignumberselected} shows that  more than a thousand hypotheses are selected, so when $\pi_1 = 0.001$ the fraction of PC nulls is very close to one (since only $50000\times 0.001=50$ PC null hypotheses  are false) but when $\pi_1 = 0.01$ the fraction of PC nulls among the selected can be far from one (since $50000\times 0.01=500$ PC hypotheses  are non-null and most of them are likely to be among the selected). 

\section{Discussion}\label{sec6}

In this paper, we introduced \CoFilter,  a powerful approach for testing multiple PC hypotheses: first select the promising candidates, which are the PC hypotheses with PC $p$-values at most a certain threshold $\tau$; then apply a valid multiple testing procedure on the conditional PC $p$-values within the selected set only. Results from simulations in  Section \ref{sec-simulations},  and data analysis in Section \ref{sec-Crohn},  highlight the potential usefulness of our approach for the discovery of consistent signals across multiple studies when the fraction of false PC null hypotheses is very small (as is often the case in GWAS).  

The main theoretical result, which provides validity for our novel method, is in Corollary \ref{cor-main}. It   holds for the important special case of point null hypotheses, so the $p$-values from true null hypotheses are uniformly distributed. We add to this the result for $p$-values from the family of specific distributions called Lehmann's alternatives in Proposition \ref{cor-LehmannAlternatives} \citep{Lehmann53}. We conjecture that  the theoretical result can be extended to more general composite nulls, for which the $p$-values from null hypotheses may be stochastically larger than uniform. An important special case of interest is the case that the null hypotheses are one-sided for parameters of exponential families.  So we would like to extend Theorem \ref{thm:expord} to the case where instead of $X_1, \ldots, X_{\ell}$ being iid $Exp(1)$ they satisfy $X_i \le_{lr} Exp(1)$. However, the proof of Theorem \ref{thm:expord} relies on a  conditioning argument that only works for Gamma random variables, so another proof is needed (as well as possibly more restrictive assumptions). 

Our theoretical guarantee is only for the Fisher combining method. An open problem is to justify other combining methods in Algorithm \ref{algo-workflow}. For example, 
using the Stouffer combination, cf. Footnote 14 in Section V of Chapter 4 of \cite{stouffer1949american},   the PC $p$-value for $H^{r/n}$ is  $1-\Phi\left(\frac{1}{\sqrt{n-r+1}}\sum_{i=r}^{n}\Phi^{-1}\left(1-p_{(i):n}\right)\right)$, where $\Phi$ and $\Phi^{-1}$ are the standard normal CDF and inverse CDF, respectively.  Another classical combining method is the one which is based on \cite{simes:1986}.  Recent combining methods are presented in \cite{Zhao2019}.

\begin{appendix}
%
\section*{Appendix: Proof of Proposition \ref{propo-BH-fixed-tau-asymptotic-FDR}}\label{appendix-lengthy-proofs}
\begin{proof}
The (random) rejection threshold for the conditional PC $p$-values which are considered in step (iii) of our Algorithm \ref{algo-workflow},  is given by
\begin{equation}\label{BH-threshold-S}
\hat t = \max\left \lbrace  t: \frac{|S_\tau|\times t \times \Diamond}{|S_{\tau t}|\lor 1}\leq \alpha\right \rbrace,
\end{equation}
where $|S_{\tau t}| = \sum_{j\in S_{\tau}} \II\left(\frac{p^{r/n}_j}{\tau}\leq t\right)= \sum_{j=1}^m \II\left(p^{r/n}_j\leq \tau t\right)$ is the number of rejected hypotheses when thresholding the conditional PC $p$-values at level $t$, and $\Diamond = 1$ for the BH procedure as well as $\Diamond = \hat{\pi}_0(S_\tau, \lambda)$ (the estimator for the proportion of true null hypotheses on the set of selected PC $p$-values) for the adaptive BH procedure. So it is enough to show that  
\begin{equation}\label{Diamond-ineq}
\lim_{m\rightarrow \infty} \sup_{0\leq t\leq 1} \left \lbrace FDP(\tau t)-\frac{|S_\tau|\times t \times \Diamond}{|S_{\tau t}|\lor 1}\right \rbrace\leq 0
\end{equation} 
almost surely, where $FDP(\tau t)$ is the false discovery proportion when the hypotheses with PC $p$-values at most $\tau t$ are rejected. 

Let $\mathcal H^{r/n} $ be the set of indices of the true PC null hypotheses (so $|\mathcal H^{r/n}| = m_0$).  
Since the conditional $p$-vales are valid, if $H_j^{r/n}$ is true, we have that $\mP\left(\frac{1}{\tau}P_j^{r/n}\leq t \right)\leq t\mP(P_j^{r/n}\leq \tau)$ for all $t\in [0,1]$. Therefore, 
\begin{eqnarray}
&& G_0(\tau t) = \lim_{m\rightarrow \infty}\frac{\sum_{j\in \mathcal H^{r/n}}\II(p_j^{r/n}\leq \tau t)}{|\mathcal H^{r/n}|}\nonumber \\
&& = \mE\left (\lim_{m\rightarrow \infty} \frac{\sum_{j\in \mathcal H^{r/n}}\II(p_j^{r/n}\leq \tau t)}{|\mathcal H^{r/n}|}\right) \nonumber \\
&& =   \lim_{m\rightarrow \infty}  \mE\left (\frac{\sum_{j\in \mathcal H^{r/n}}\II(p_j^{r/n}\leq \tau t)}{|\mathcal H^{r/n}|}\right) =  \lim_{m\rightarrow \infty} \frac{\sum_{j\in \mathcal H^{r/n}}\mP(p_j^{r/n} \leq \tau t)}{|\mathcal H^{r/n}|}
 \nonumber \\ 
&&
\leq \lim_{m\rightarrow \infty} \frac{\sum_{j\in \mathcal H^{r/n}}\mP(p_j^{r/n}\leq \tau )t}{|\mathcal H^{r/n}|} = G_0(\tau) t.  \label{eq-asymp1}
\end{eqnarray}

From (a slight modification) of the Glivenko-Cantelli theorem \citep{Storey03}, 
$$\lim_{m\rightarrow \infty} \sup_{0\leq t\leq 1}\left|\frac{|S_t|}{m}-\pi_0G_0(t)-(1-\pi_0)G_1(t) \right|=0$$ 

almost surely. Therefore, 
\begin{eqnarray}\label{eq-asymp2}
\lim_{m\rightarrow\infty } \sup_{0\leq t\leq 1} \left |\frac{|S_\tau|\times t}{|S_{\tau t}|\lor 1} - \frac{\pi_0G_0(\tau)t+(1-\pi_0)G_1(\tau)t}{(\pi_0G_0(\tau t)+(1-\pi_0)G_1(\tau t))\lor m^{-1}}\right|= 0
\end{eqnarray} 
almost surely.

We now use \eqref{eq-asymp1} and \eqref{eq-asymp2} in order to show that \eqref{Diamond-ineq} holds true both for the BH procedure and for the adaptive BH procedure. We start by expressing the difference within the curly brackets in \eqref{Diamond-ineq} as the sum of three differences, and then we argue  that  each of these differences is almost surely upper bounded by zero for eventually all large $m$. Namely, 
\begin{eqnarray}
&& FDP(\tau t)-\frac{|S_{\tau}| \times t \times \Diamond}{|S_{\tau t}|\lor 1} =\frac{\sum_{j\in \mathcal H^{r/n}}\II(p_j^{r/n}\leq \tau t)}{\sum_{j=1}^m\II(p_j^{r/n}\leq \tau t)\lor 1} -\frac{|S_{\tau}| \times t \times \Diamond}{|S_{\tau t}|\lor 1}\nonumber \\
&& =  \frac{\sum_{j\in \mathcal H^{r/n}}\II(p_j^{r/n}\leq \tau t)}{\sum_{j=1}^m\II(p_j^{r/n}\leq \tau t)} -\frac{\pi_0G_0(\tau t)}{\pi_0G_0(\tau t) +(1-\pi_0) G_1(\tau t)} \nonumber \\
&& +   \frac{\pi_0G_0(\tau t)}{\pi_0G_0(\tau t) +(1-\pi_0) G_1(\tau t)} - \frac{\pi_0G_0(\tau)t}{\pi_0G_0(\tau t)  +(1-\pi_0) G_1(\tau t)}   \nonumber \\
&&+ \frac{\pi_0G_0(\tau)t}{\pi_0G_0(\tau t)  +(1-\pi_0) G_1(\tau t)} - \frac{|S_{\tau}| \times t \times \Diamond}{|S_{\tau t}|\lor 1}.  \label{third-difference}
\end{eqnarray}

Now, from assumptions \eqref{eq-conv1}--\eqref{eq-conv3}, as in \cite{Storey03}, 
$$
\lim_{m\rightarrow\infty } \sup_{0\leq t\leq 1} \left | \frac{\sum_{j\in \mathcal H^{r/n}}\II(p_j^{r/n}\leq \tau t)}{(\sum_{j=1}^m\II(p_j^{r/n}\leq \tau t))\lor 1} -\frac{\pi_0G_0(\tau t)}{(\pi_0G_0(\tau t) +(1-\pi_0) G_1(\tau t))\lor m^{-1}}\right| = 0
$$
holds almost surely, and from \eqref{eq-asymp1}, 
$$
\frac{\pi_0G_0(\tau t)}{(\pi_0G_0(\tau t) +(1-\pi_0) G_1(\tau t))\lor m^{-1}} - \frac{\pi_0G_0(\tau)t}{(\pi_0G_0(\tau t)  +(1-\pi_0) G_1(\tau t))\lor m^{-1}}\leq 0.
$$
Thus, it remains to discuss the third difference in \eqref{third-difference}. We do this for both values of $\Diamond$ separately.\\

\underline{Case (i), $\Diamond = 1$:}\\
In this case, we immediately get from \eqref{eq-asymp2} that 
$$
\lim_{m\rightarrow\infty } \sup_{0\leq t\leq 1} \left \lbrace  \frac{\pi_0G_0(\tau)t}{(\pi_0G_0(\tau t)  +(1-\pi_0) G_1(\tau t))\lor m^{-1}} - \frac{|S_{\tau}| \times t}{|S_{\tau t}|\lor 1} \right \rbrace \leq 0.
$$

\underline{Case (ii), $\Diamond = \hat{\pi}_0(S_\tau, \lambda)$:}\\
In this case, we first notice that
\[
\hat{\pi}_0(S_\tau, \lambda) = \frac{|S_\tau| - |S_{\tau \lambda}|}{(1 - \lambda) |S_\tau|}.
\]
With this representation, the third difference in \eqref{third-difference} becomes
\[
\frac{\pi_0G_0(\tau)t}{\pi_0G_0(\tau t)  +(1-\pi_0) G_1(\tau t)} - \frac{|S_{\tau}| \times t \times \{|S_\tau| - |S_{\tau \lambda}|\}}{\{|S_{\tau t}|\lor 1\} \times (1 - \lambda) \times |S_\tau|},       
\]
which is upper-bounded by zero if 
\[
\frac{\pi_0G_0(\tau)}{\pi_0G_0(\tau t)  +(1-\pi_0) G_1(\tau t)} \leq \frac{\{|S_\tau| - |S_{\tau \lambda}|\} \times m^{-1}}{\{|S_{\tau t}|\lor 1\} \times (1 - \lambda) \times m^{-1}}.
\]
Since $\{|S_{\tau t}|\lor 1\} / m$ converges almost surely to $\pi_0G_0(\tau t)  +(1-\pi_0) G_1(\tau t)$ as $m \to \infty$, it suffices to analyze whether 
\begin{equation}\label{ruth-two-stars}
\pi_0G_0(\tau) \leq \frac{|S_\tau| - |S_{\tau \lambda}|}{m (1 - \lambda)}  
\end{equation}
holds true for eventually all large $m$. To verify \eqref{ruth-two-stars}, we notice that
\begin{eqnarray*}
\frac{|S_\tau| - |S_{\tau \lambda}|}{m (1 - \lambda)} &= &\frac{\sum_{j\in \mathcal H^{r/n}}\II(\lambda < \frac{1}{\tau} p_j^{r/n}\leq 1) + \sum_{j \not \in \mathcal H^{r/n}}\II(\lambda < \frac{1}{\tau} p_j^{r/n}\leq 1)}{m (1 - \lambda)}\\
&\geq &\frac{\sum_{j\in \mathcal H^{r/n}}\II(\lambda < \frac{1}{\tau} p_j^{r/n}\leq 1)}{m (1 - \lambda)}.
\end{eqnarray*}
Thus, it suffices to show that 
\begin{equation}\label{ruth-preultimate}
\lim_{m \to \infty} \frac{|H^{r/n}|}{m} \frac{\sum_{j\in \mathcal H^{r/n}}\II(\lambda < \frac{1}{\tau} p_j^{r/n}\leq 1)}{(1 - \lambda) |H^{r/n}|} \geq \pi_0 G_0(\tau)
\end{equation}
holds true almost surely. To see this, notice that
$\lim_{m \to \infty} \frac{|H^{r/n}|}{m} = \pi_0$,  and that for each $j \in H^{r/n}$ we have from Corollary \ref{cor-main} that $\mathbb{P}(\frac{1}{\tau} p_j^{r/n} > \lambda  | p_j^{r/n}\leq \tau) \geq 1-\lambda$, implying that $\mathbb{P}(\lambda < \frac{1}{\tau} p_j^{r/n}\leq 1) \geq (1-\lambda) \mathbb{P}( p_j^{r/n}\leq \tau)$. By virtue of \eqref{eq-conv1}, this implies that \eqref{ruth-preultimate} holds true almost surely, completing the proof.
\end{proof}
\end{appendix}


\section*{Acknowledgement}
We are indebted to Yaming Yu for his essential contribution to the proof of Theorem \ref{thm:expord}.

\begin{funding}
The authors gratefully acknowledge financial support by the German Research Foundation via Grant No. DI 1723/5.  YR was supported in part by a grant from the Center for Interdisciplinary Data Science Research at the Hebrew University (CIDR). RH was supported in part by Israeli Science Foundation grant
no. 2180/20.
\end{funding}


\end{document}